\documentclass{article}
\usepackage{latexsym}

\begin{document}

\def\EQN#1{eq.(\ref{eq:#1})}
\def\half{\!\matrix{\frac{1}{2}}\!}
\def\fraction#1#2{\!\matrix{\frac{#1}{#2}}\!}
\def\C{{\bf C}}
\def\gammahat{{\hat \gamma}}
\def\phihat{{\gamma}}
\def\bo{\beta_\mathrm{o}}
\def\be{\beta_\mathrm{e}}
\def\co{\chi_\mathrm{o}}
\def\ce{\chi_\mathrm{e}}
\def\bev{b_\mathrm{e}}
\def\cev{c_\mathrm{e}}
\def\bod{b_\mathrm{o}}
\def\cod{c_\mathrm{o}}
\def\xe{x_\mathrm{e}}
\def\ye{y_\mathrm{e}}
\def\yo{y_\mathrm{o}}
\def\xo{x_\mathrm{o}}
\def\Xo{X_\mathrm{o}}
\def\k{\kappa}

\def\spaces{\ \ \ \ \ \ \ \ \ \ }

\def\bog#1#2{\beta_{#1}^{\mathrm{o}#2}}
\def\beg#1#2{\beta_{#1}^{\mathrm{e}#2}}
\def\cog#1#2{\chi_{#1}^{\mathrm{o}#2}}
\def\ceg#1#2{\chi_{#1}^{\mathrm{e}#2}}
\def\e{\mathrm{e}}
\def\o{\mathrm{o}}
\def\sech{\mathrm{sech}}
\def\manytabs{&\ &\ \ \ \ \ \ \ \ \ \
\ \ \ \ \ \ \ \ \ \ \ \ \ \ \ \ \ \ \ \ \ \ \ \ \ \ \ \ \ \ \ \ \
\ \ \ \ \ \ \ \ \ \ \ \ \ \ \ \ \ \ \ \ \ \ \ \ \ \ \ \ \ \ \ \ }
\def\Bo#1{b^{\mathrm{o}#1}}
\def\Be#1{b^{\mathrm{e}#1}}
\def\Co#1{c^{\mathrm{o}#1}}
\def\Ce#1{c^{\mathrm{e}#1}}
\def\Xos#1{x^{\mathrm{o}#1}}
\def\Ye#1{y^{\mathrm{e}#1}}
\def\Xe#1{x^{\mathrm{e}#1}}
\def\Yo#1{y^{\mathrm{o}#1}}
\def\dd#1{\partial_{#1}}
\def\pr{\mathrm{'}}

\def\der#1{\frac{\partial}{\partial #1}}
\def\AUTHORA{Theodore G. Erler}
\def\EMAILA{terler@physics.ucsb.edu}
\def\TITLE{Moyal Formulation of Witten's Star
Product in the Fermionic Ghost Sector}

\begin{titlepage}

\title{\Large\bf \TITLE}
\author{{\large  \AUTHORA}
\thanks{Email:\EMAILA} \\
\\ Department of Physics
\\ University of California
\\ Santa Barbara, CA}
\maketitle \thispagestyle{empty}

\begin{abstract}

In this paper, we recast the fermionic ghost sector of Witten's
open bosonic string field theory in the language of noncommutative
field theory. In particular, following the methods of
hep-th/0202087, we find that in Siegel gauge Witten's star product
roughly corresponds to a continuous tensor product of Clifford
Algebras, and we formulate important operators of the theory in
this language, notably the kinetic operator of vacuum string field
theory and the BRST operator describing the vacuum of the unstable
D-25 brane. We find that the BRST operator is singular in this
formulation. We explore alternative operator/Moyal representations
of the star product analogous to the split string description and
the discrete Moyal basis developed extensively in recent work by
Bars and Matsuo (hep-th/0204260). Finally, we discuss some
interesting singularities in the formalism and how they may be
regulated.

\end{abstract}

\end{titlepage}
\section*{I. Introduction}

Since Sen described his famous conjectures on tachyon
condensation\cite{Sen-conjectures}, there has been renewed
interest in the framework of string field theory\cite{Witten},
both as a tool for studying tachyon condensation and as a possible
nonperturbative and background independent formulation of string
theory. To this day, most of our knowledge of the theory has come
about through more or less brute force numerical analysis of the
string field equations\cite{Kostelecky}---level truncation. Some
analytic insight however has emerged through the study of vacuum
string field theory (VSFT), a theory constructed so as to
supposedly describe physics around the locally stable closed
string vacuum, and which possesses a particularly simple kinetic
operator, $\mathcal{Q}=c^+(\frac{\pi}{2} )+c^-(\frac{\pi}{2})$
\cite{VSFT, Sen}. However, it remains unclear whether VSFT is a
nonsingular or even accurate description of string theory---around
the closed string vacuum or otherwise. We need more powerful
analytic techniques, particularly to study the full open string
field theory, whose kinetic operator, the BRST operator, is much
more complicated.

Other than the usual oscillator and conformal field theory
methods\cite{Gross-Jevicki-1, Gross-Jevicki-2, LeClair}, three
alternative approaches have been recently proposed for studying
string field theory in a possibly simpler framework: the split
string formalism\cite{half-strings, Gross-Taylor-2,
Kuwano-Okuyama, Bordes}, the discrete Moyal formalism advocated by
Bars and Matsuo\cite{Bars,Bars-Matsuo}, and the continuous Moyal
formalism\cite{Moyal, Aref'eva, Belov, Chen}. All of these
approaches have a common goal, which is to reformulate string
field theory in the language of operator algebras and
noncommutative geometry\cite{Douglas,Aref'eva_review}. Witten's
star product is then simple, roughly corresponding to matrix
multiplication or noncommutative Moyal star multiplication. Of
course, these two descriptions are naively isomorphic. This
represents a marked improvement over the usual oscillator
approach, where the star product is notoriously complicated.
However, the down side of these new approaches is that the BRST
operator seems very difficult to formulate, and may in fact be
ill-defined. Worse, in the split string formalism it does not even
seem possible to accommodate operators which act nontrivially on
the string midpoint, such as the pure ghost $\mathcal{Q}$ of VSFT.

Of course, a clear understanding of these issues requires an
adequate formulation of the ghost sector of the theory. For the
most part, the ghost sector has been avoided for the simple reason
that it is unclear how Witten's product in the ghost sector should
be described as either a matrix or Moyal product. In references
\cite{Gross-Taylor-2,Bars-Matsuo} the approach has been to
bosonize the ghosts, in which case the star product is given by a
simple half-string overlap (times a phase factor inserting ghost
momentum at the midpoint) which makes an operator/Moyal
description seem natural. However, in the bosonized language the
BRST operator is extremely complicated, and it becomes a labor
even to verify the axioms of string field theory for the extremely
simple pure ghost kinetic operators studied in
ref.\cite{Gross-Taylor-2}. Furthermore, no nontrivial solution to
VSFT has been satisfactorily constructed using the bosonized
ghosts. Therefore it would seem extremely advantageous to find an
operator/Moyal formulation of string field theory using the
ordinary fermionic ghosts. It is known, for instance, that star
multiplication in the fermionic ghost sector roughly corresponds
to a half string overlap in the $b$ ghost and an {\it anti}overlap
(overlap up to a sign) in the $c$ ghost. This suggests that it may
be possible to write the fermionic star product as a matrix
product but with an additional operator insertion to impose the
requisite anti-overlap conditions on the $c$ ghost:
\begin{equation} \Psi*\Phi
\sim\hat{\Psi}\hat{I}^{-1}\hat{\Phi}\label{eq:ferm-split-string}
\end{equation} The half string operator $\hat{I}^{-1}$
should be the inverse of the identity string field\footnote{Note
that the identity string field serves as the identity of Witten's
star product, {\it not} the matrix product of half string
operators. This is why the insertion $\hat{I}^{-1}$ in
\EQN{ferm-split-string} is nontrivial}. However, if one pursues
this avenue, one finds that the identity string field is roughly,
\begin{equation}\hat{I}=N\bar{b}\int[d\lambda][d\mu]|\lambda,\mu\rangle
\langle \lambda,-\mu|\end{equation} where $N$ is a normalization
and,
$$\bar{b}=b(\fraction{\pi}{2}),\ \
\lambda(\sigma)=-\frac{2}{\pi}\sigma\bar{b}+b(\sigma),\ \
\mu(\sigma)=c(\sigma)$$ for $\sigma\in[0,\frac{\pi}{2}]$. The
problem is that $\hat{I}$ is not invertable! The $\bar{b}$
multiplying the integral clearly has no inverse, and comes from
necessary midpoint insertions\cite{Gross-Jevicki-2,Samuel} giving
the identity string field the correct ghost number. This is
somewhat baffling since in the bosonized ghost language $\hat{I}$
is a perfectly invertible number, $M^{-1}=e^{-i3\bar{\phi}/2}$,
where $\bar{\phi}$ is the midpoint value of the bosonized ghost.
Nevertheless, it is clearly impossible to make sense of
\EQN{ferm-split-string} given an identity field which is not
invertible, so it is very unclear how the fermionic star product
may be realized as any sort of matrix or Moyal product.

However recent studies of the spectrum of eigenvalues of the
Neumann coefficients\cite{Spectroscopy} have made it possible to
choose a (continuous) basis of oscillators which diagonalize the
quadratic form appearing in zero momentum three string
vertex\cite{Moyal}. The authors of \cite{Moyal} discovered that
the diagonalized vertex exactly describes a continuous tensor
product of Heisenberg algebras, and were able to make a direct
connection with the split string and discrete Moyal formalisms
developed earlier. The remarkable thing about their approach is
that it can be extended straightforwardly to the fermionic ghost
sector, since in Siegel gauge the three string ghost vertex may be
diagonalized by more or less the same techniques. In this way we
may hope to recover an operator/Moyal formulation of the fermionic
ghost star product even though other attempts have
failed\cite{Gross-Taylor-2, Kuwano-Okuyama}.

This is the subject of the current paper. We find that, in Siegel
gauge, the so-called reduced star product\cite{Okuyama2} ($=b_0$
times the full star product) corresponds to a continuous tensor
product, from $\k =0$ to $\infty$, of a pair of Clifford Algebras
$Cl_{1,1}$, the first with a metric scaling as $\coth\frac{\pi \k
}{4}$ and the second with a metric scaling as $-\coth\frac{\pi \k
}{4}$. Clifford Algebra, in a sense we will explain, can be
thought of as defining non-anticommutative geometry on a Grassmann
space. Therefore our results show that string field theory may
indeed be formulated in the language of noncommutative field
theory. We take advantage of this and formulate both the kinetic
operator of VSFT and the BRST operator in this new basis. The BRST
operator in particular is divergent, and must somehow be regulated
to get a consistent formulation. We use our results in the
continuous Moyal formalism to construct split string and discrete
Moyal representations of the reduced star product in Siegel gauge.
We hope these results encourage new progress in formulating string
field theory in a language where Witten's product is simple.

As a matter of notation, we will denote the Moyal product with a
five pointed star,  $\star$, Witten's product with a six pointed
one $*$, and the reduced star product with $*_{b_0}$.

\section*{II. Non-anticommutative geometry on
Grassmann spaces} In this section we would like to develop the
notion of non(anti)commutative geometry on a space with Grassmann
valued coordinates\footnote{Related discussion can be found in
\cite{Aref'eva,Ferrara}.}. Let us start with familiar notions from
noncommutative geometry on ordinary flat space. Given any vector
space $\Re^{2n}$ with a nondegenerate 2-form $\theta$, we can
construct an associative, noncommutative algebra
$\mathcal{A}_\theta$ defined as the quotient:\begin{equation}
\mathcal{A}_\theta\equiv\frac{\C\otimes\mathrm{Tensor}(\Re^{2n})}{\{[{\bf
u},{\bf v}]\sim i\theta({\bf u},{\bf v}),\ \  {\bf u},{\bf v}\in
\Re^{2n}\}}\label{eq:A} \end{equation} Where
$\mathrm{Tensor}(\Re^{2n})$ denotes the tensor algebra over
$\Re^{2n}$. Intuitively, $\mathcal{A}_\theta$ is the algebra
generated by taking all products and complex linear combinations
of $2n$ basis vectors ${\bf x}_i$ for $\Re^{2n}$ subject to the
multiplication rule $[{\bf x}_j,{\bf x}_k]=i\theta_{jk}$. We
should think of $\mathcal{A}_\theta$ as a noncommutative
deformation of the algebra\footnote{The algebra of continuous
functions is, naturally, the algebra defined with the addition and
multiplication laws $(f\cdot g)(x)=f(x)g(x)$ and
$(f+g)(x)=f(x)+g(x)$.} of continuous complex functions on
$\Re^{2n}$. In this sense $\mathcal{A}_\theta$ defines what we
mean by {\it noncommutative geometry} on $\Re^{2n}$. Since
$\theta$ is nondegenerate, we can always choose a basis so that
$\theta_{ij}$ is block diagonal, and in this basis we have:
\begin{equation}
[{\bf x}_{2j-1},{\bf x}_{2j}]=i,\ \ j=1,...,n,
\end{equation}
with all other commutators vanishing. With simple renaming of
variables, ${\bf \hat{x}}_i={\bf x}_{2i-1}$ and ${\bf
\hat{p}}_i={\bf x}_{2i}$, the commutation relations read: $$[{\bf
\hat{x}}_j,{\bf \hat{p}}_k]=i\delta_{jk},\ \ \ [{\bf
\hat{x}}_j,{\bf \hat{x}}_k]=0,\ \ \ [{\bf \hat{p}}_j,{\bf
\hat{p}}_k]=0.$$ This, of course, looks just like the canonical
commutation relations for the position and momentum operators
describing the quantum mechanics of a point particle moving in $n$
spatial dimensions. To make this correspondence exact, however, we
must define a Hilbert space so that the ${\bf x}_i$'s act as
Hermitian operators. This defines a representation of the algebra
$\mathcal{A}_\theta$, which, in turn, induces a natural {\it
involution} on $\mathcal{A}_\theta$ corresponding to Hermitian
conjugation.

Our discussion suggests that we might define non-anticommutative
geometry on a {\it Grassmann space} by following the above
argument in reverse. In particular, start with the quantum
mechanics of a particle with Grassmann-valued coordinates; then we
have hermitian position and momentum operators ${\hat \xi}_i,\
{\hat \pi}_i$ $i=1,...,n$ satisfying canonical anticommutation
relations: $$\{{\hat \xi}_j,{\hat \pi}_k\}=\delta_{jk},\ \ \
\{{\hat \xi}_j,{\hat \xi}_k\}=0,\ \ \ \{{\hat \pi}_j,{\hat
\pi}_k\}=0.$$ Now simply rename ${\hat \xi}_i=\gamma_{2i-1}$ and
${\hat \pi}_i=\gamma_{2i}$, then we get:
$$\{\gamma_{2j-1},\gamma_{2j}\}=1,$$ with all other
anticommutators vanishing. Through a (real) linear change of
basis, $\gamma_i\to M_{ij}\gamma_j$ with $M=M^*$ this relation can
be put in the general form:
\begin{equation}\{\gamma_i,\gamma_j\}=2g_{ij}\label{eq:Clifford}
\end{equation} with $$g=\half MfM^T,\ \ \ \ f=
\pmatrix{\matrix{0 & 1\cr 1 & 0} & \ & \ \cr \ & \ddots & \ \cr \
& \ & \matrix{0 & 1 \cr 1 & 0}}.$$ We instantly recognize
\EQN{Clifford} as the defining relation for a Clifford Algebra.
Note in particular that we can diagonalize $f$ through a suitable
choice of $M$:
$$M=\pmatrix{\matrix{1
& -1 \cr 1 & \ 1} & \ & \ \cr \ & \ddots & \ \cr \ & \ & \matrix{1
& -1 \cr 1 & \ 1}},\ \ \ \Longrightarrow\ \ \
g=\pmatrix{\matrix{-1 & 0 \cr \ 0 & 1} & \ & \ \cr \ & \ddots & \
\cr \ & \ & \matrix{-1 & 0 \cr \ 0 & 1}}.$$ Therefore,
\EQN{Clifford} defines a Clifford Algebra with respect to a metric
of signature $(n,n)$. Putting this all together, we define an
associative algebra:
\begin{equation}\mathcal{B}_g=\C \otimes
Cl_{n,n}=\frac{\C\otimes\mathrm{Tensor}(\Re^{2n})}{\{{\bf v}^2\sim
g({\bf v},{\bf v}),\ \  {\bf v}\in
\Re^{2n}\}},\label{eq:algebra}\end{equation} which is a
non-anticommutative deformation of the (Grassmann) algebra of
complex functions on a $2n$ dimensional Grassmann space. This
algebra defines what we mean by non-anticommutative geometry on a
Grassmann space.

An important point: Since we derived $\mathcal{B}_g$ from the
quantum mechanics of a fermionic point particle, the algebra
inherits a natural involution corresponding to Hermitian
conjugation. This allows us to distinguish between the algebras
$\C\otimes Cl_{n,n}$ and $\C\otimes Cl_{k,2n-k}$, which after all
only differ by absorbing a factor of $i$ in the choice of the
basis. However, this type of redefinition is not allowed, since a
new basis obtained this way would not be hermitian. This is why we
required $M=M^*$ in \EQN{Clifford}. This, however, raises the
question of whether we should generalize our definition of
non-anticommutative geometry by setting $\mathcal{B}_g=\C\otimes
Cl_{p,q}$ for arbitrary metric signature $(p,q)$. The problem is
that these more general algebras will not have a simple
representation in terms of the quantum mechanics of a fermionic
point particle. For the case of ordinary noncommutative geometry,
the correspondence to the quantum mechanics of a point particle is
automatic, regardless of our choice of $\theta$. This is not true
for the fermionic case regardless of our choice of $g$. For the
purposes of this paper, the narrower definition \EQN{algebra} of
$\mathcal{B}_g$ is the most appropriate.

Often in noncommutative geometry we represent the algebra
$\mathcal{A}_\theta$ as an algebra of continuous functions on
$\Re^{2n}$ which are multiplied together with the noncommutative
and nonlocal Moyal star product. We can similarly find a
representation of $\mathcal{B}_g$ in terms of a nonlocal product
between functions on a Grassmann space (whose coordinates we write
as $\phi_i, i=1...2n$),
\begin{equation}f\star
g(\phi)=\left.\mathrm{exp}\left[g_{ij}\der{\phi_i}
\der{\phi'_j}\right]f(\phi')g(\phi)\right|_{\phi=\phi'}.\label{eq:Moyal}
\end{equation} The relationship between this analytic
representation of $\mathcal{B}_g$ and the algebraic representation
described in the last paragraph is given precisely as follows.
Given a function $f(\psi)$ on our Grassmann space, we can
construct an antisymmetrically ordered operator $O_f$ as follows,
$$O_f\equiv\int
d^{2n}\phi \tilde{f}(\phi) e^{-\phi^i\gamma_i},\ \ \
\tilde{f}(\phi)=\int d^{2n}\psi f(\psi)e^{\phi^i\psi_i},$$ where
the operators $\gamma_i$ satisfy \EQN{Clifford}. Then we can
write: $$O_f O_g=O_{f\star g}.$$ The operator and Moyal
representations of the algebra are isomorphic.

Let's focus on the most elementary case of a two dimensional
Grassmann space with coordinates $x$ and $y$ and metric
$g_{ij}=g\pmatrix{0 & 1\cr 1 & 0}_{ij}$. For short we will denote
$X=(x,y)$. For the purposes of this paper we will find it
extremely useful to represent non-anticommutative multiplication
in terms of a fermionic integral kernel,
\begin{equation}f\star g(X^3)=\int
\prod_{B=1,2}dx^B dy^B
K(X^1,X^2,X^3)f(X^1)g(X^2),\label{eq:kernel-form}\end{equation}
From \EQN{Moyal} we know that, $$f\star g(X^3)=\int dx^1 dy^1 dx^2
dy^2 (y^2-y^3)(x^2-x^3) (y^1-y^3)(x^1-x^3)
e^{g(\partial^2_x\partial^1_y
+\partial^2_y\partial^1_x)}f(X^1)g(X^2).$$ Integrating by parts,
we can take the derivatives off of $f$ and $g$ leading to an
expression for the kernel:\begin{eqnarray}K(X^1,X^2,X^3)&=&
e^{g(\partial^2_x\partial^1_y +\partial^2_y\partial^1_x)}
(y^2-y^3)(x^2-x^3) (y^1-y^3)(x^1-x^3)\nonumber\\ &=&(y^2-y^3)
(x^2-x^3) (y^1-y^3)(x^1-x^3)- g(y^2-y^3)(x^1-x^3)\nonumber\\ &\ &\
\ \ -g(x^2-x^3)(y^1-y^3)+g^2\nonumber\\ &=&g^2
\mathrm{exp}\left[g^{-1}(x^1y^2+x^2y^3
+x^3y^1-x^2y^1-x^3y^2-x^1y^3)\right]\nonumber\end{eqnarray} More
succinctly,
\begin{equation}K(X^1,X^2,X^3)=g^2e^{x^A K^{AB}
y^B},\ \ \ \ \ K^{AB}=g^{-1}\pmatrix{0 & 1 & -1 \cr -1 & 0 & 1 \cr
1 & -1 & 0}^{AB},\label{eq:kernel}\end{equation} with an implied
summation over the indices $A,B=1,2,3$. The kernel is invariant
under cyclic permutations of $1,2,3$. An important property of
this kernel is,
$$(\partial^1+\partial^2+\partial^3)K(X^1,X^2,X^3)=0$$
where $\partial$ represents a derivative with respect to either
$x$ or $y$. This statement is equivalent to the nontrivial fact
that $\partial$ acts as a derivation of the $\star$ algebra,
$$\partial(f\star g)=(\partial f)\star g+(-1)^{G(f)}
f\star\partial g $$ where $G(f)$ is the Grassmann rank of $f$
(assuming it has a definite rank). This allows us to represent the
action of the derivative on the algebra as a star
(anti)commutator, just as in ordinary noncommutative field theory:
\begin{equation}\partial_x
f(x,y)=\frac{1}{2g}\left(y\star f(x,y)-(-1)^{G(f)}f(x,y)\star y
\right)\label{eq:derivative-commutator}\end{equation} with a
similar relation for $\partial_y$. To see that these two
operations are equivalent, all we have to do is check that they
act identically on $1,x$, and $y$ and it follows that they will
act identically on $f(x,y)$ from linearity and the above derived
product rule.

\section*{III. Diagonalizing the fermionic ghost
vertex} We will now simplify the three-string vertex in the
fermionic ghost sector, along similar lines as ref.\cite{Moyal},
by choosing a basis of operators which diagonalize the quadratic
form appearing in the vertex. In this paper we use the same
conventions for the fermionic ghosts as
\cite{GSW,Gross-Jevicki-2}. The three string vertex in the
fermionic ghost sector was constructed in
ref.\cite{Gross-Jevicki-2,Samuel}:
\begin{equation}|V_3\rangle =
\mathrm{exp}\left[-b_0^A\tilde{N}_{0m}^{AB}\sqrt{m}c^{B+}_m-
b^{A+}_m\frac{1}{\sqrt{m}}\tilde{N}^{AB}_{mn}\sqrt{n}c^{B+}_n\right]|+\rangle_1
|+\rangle_2|+\rangle_3.\label{eq:vertex}\end{equation} Here, $A,B$
are Hilbert space indices ranging from one to three, and $m,n$ are
level number indices ranging from $1$ to $\infty$, repeated
indices summed. Given two string fields $|\phi\rangle$,
$|\psi\rangle$ we calculate\footnote{For simplicity we will assume
our string fields are real, so \EQN{star} can be taken as the
definition of the star product. For fields which are not real, the
left hand side of \EQN{star} should be $\langle
\phi*\psi||V_2\rangle$ where $|V_2\rangle$ is the two string
vertex.} their star product by evaluating,
\begin{equation}|\phi*\psi\rangle=\langle\phi
|\langle\psi||V_3\rangle\label{eq:star}\end{equation} We will find
it useful to restrict ourselves to evaluating the star product in
Siegel gauge. String fields will then be of the form
$|\psi\rangle=|\hat{\psi}\rangle\otimes|-\rangle$ where
$|\hat{\psi}\rangle$ is in the Fock space generated by the level
$n>1$ $bc$-oscillators. Multiplying \EQN{star} by $b_0$, we obtain
a formula for the so-called ``reduced''\cite{Okuyama2} star
product in Siegel gauge: \begin{equation} |\phi*_{b_0} \psi\rangle
\equiv b_0|\phi*\psi\rangle=|-\rangle\otimes\langle \hat{\phi}|
\langle \hat{\psi}||\hat{V}_3\rangle,
\label{eq:reducedstar}\end{equation} where,
\begin{equation}|\hat{V}_3\rangle =
\mathrm{exp}\left[-b^{A+}_m\frac{1}{\sqrt{m}}\tilde{N}^{AB}_{mn}\sqrt{n}
c^{B+}_n\right]|0\rangle_1|0\rangle_2
|0\rangle_3.\label{eq:vertex}\end{equation} with $|0\rangle$ being
the vacuum annihilated by $b_n,c_n$ for $n\geq 1$. As explained in
ref.\cite{Okuyama2}, if we know the reduced star product of two
string fields in Siegel gauge, at least formally we can calculate
their full star product using the relation:
\begin{equation}|\phi*\psi\rangle
=\mathcal{Q}|\phi*_{b_0}\psi\rangle \label{eq:fullstar}
\end{equation} where
$\mathcal{Q}=c^+(\fraction{\pi}{2})+c^-(\fraction{\pi}{2})$ is the
canonical choice of pure ghost kinetic operator in vacuum string
field theory, as discussed in \cite{Hata,Sen,Okuyama}. Equation
(\ref{eq:fullstar}) follows from the observation of
ref.\cite{Okuyama2} that the star product of two fields in Siegel
gauge is formally $\mathcal{Q}$ exact.

To better understand the structure of the star product for the
fermionic ghosts, we can therefore gauge fix and focus on the
vertex for the reduced star product, $|\hat{V}_3\rangle$.
Following \cite{Moyal}, we will simplify this vertex by choosing a
particular basis of oscillators in which the Neumann coefficients
$\tilde{N}^{AB}$ are diagonal. To do this, however, we must
understand the spectrum of eigenvectors and eigenvalues of
$\tilde{N}^{AB}$. Fortunately, this is not too difficult, since
the $\tilde{N}^{AB}$ are closely related to the zero momentum
Neumann coefficients $N^{AB}$ appearing in the matter sector
vertex\cite{Gross-Jevicki-1} whose eigenvectors and eigenvalues
have been explicitly calculated in \cite{Spectroscopy}.

It is convenient to define matrices $M^{AB}\equiv CN^{AB}$
($C_{mn}=(-1)^m\delta_{mn}$) which satisfy the important
relations\cite{Hata}: $$[M^{AB},M^{A'B'}]=0,\ \ \
(M^{AB})^+=M^{AB},\ \ M^{AB}=M^{A+1,B+1}$$
\begin{equation}M+M^{12}+M^{21}=1,\ \
M^{12}M^{21}=M(M-1).\label{eq:M-relations}\end{equation} These
relations imply that there are three independent Neumann matrices,
$M^{12}$, $M^{21}$, and $M^{11}=M$ which can all be simultaneously
diagonalized. Similar relations hold for the ghost matrices
$\tilde{M}^{AB}\equiv C\tilde{N}^{AB}$. In fact, we can write the
ghost matrices $\tilde{M}^{AB}$ explicitly in terms of the
matrices $M^{AB}$ as follows\footnote{Our expressions for the
Neumann matrices differ from those in ref.\cite{Belov}.}:
\begin{equation}\tilde{M}^{12}=\frac{1+M-M^{21}}{1+2M},\
\ \ \ \tilde{M}^{21}=\frac{1+M-M^{12}}{1+2M}\ \ \ \
\tilde{M}=-\frac{M}{1+2M}.\label{eq:M-tilde}\end{equation} These
nonlinear relations are different from the usual formula defining
$\tilde{M}^{AB}$ in terms of six string Neumann
coefficients\cite{Gross-Jevicki-2}, but they can be implied from
the expressions in ref.\cite{Gross-Jevicki-2} or derived
independently from the Moyal formalism developed extensively in
ref.\cite{Bars-Matsuo} (in particular, from equations 5.35 and
5.47 of that reference). It is an easy check to verify that the
ghost matrices in \EQN{M-tilde} satisfy \EQN{M-relations}. Clearly
the eigenvectors of $M^{AB}$ will also be eigenvectors of
$\tilde{M}^{AB}$.

So let us recall the basic results of \cite{Spectroscopy,Okuyama}.
The matrices $M^{AB}$ have a continuous spectrum of eigenvectors
$v_n(\k)$ for $-\infty<\k<\infty$:
\begin{equation}M^{AB}_{mn}v_n(\k)=\mu^{AB}(\k )v_m(\k),\label{eq:evectors}\end{equation}
with,\begin{eqnarray} \mu(\k) &=
&-\frac{1}{1+2\mathrm{cosh}\frac{\pi \k}{2}}, \nonumber\\
\mu^{12}(\k) &= &\frac{1+\mathrm{cosh}\frac{\pi
\k}{2}+\mathrm{sinh}\frac{\pi \k}{2}}{1+2\mathrm{cosh}\frac{\pi
\k}{2}}, \nonumber\\ \mu^{21}(\k) &=
&\frac{1+\mathrm{cosh}\frac{\pi \k}{2}-\mathrm{sinh}\frac{\pi
\k}{2}}{1+2\mathrm{cosh}\frac{\pi
\k}{2}}.\label{eq:evalues}\end{eqnarray} The special functions
$v_n(\k)$ can be derived from the generating functional:
\begin{equation}\sum_{n=1}^\infty
\frac{v_n(\k)}{\sqrt{n}}z^n=\frac{1-e^{-\k\mathrm{tan}^{-1}z
}}{\k\sqrt{\frac{2}{\k}\mathrm{sinh}\frac{\pi
\k}{2}}}\label{eq:genr-f}\end{equation} The normalization factor
in the denominator is an odd function of $\k$. The functions
$v_n(\k)$ enjoy the following properties:
\begin{eqnarray}(-1)^{n+1}v_n(\k)&=&v_n(-\k),\nonumber\\
v_n(\k)v_n(\k')=\delta(\k-\k'),\ &\ &\ \
\int_{-\infty}^{\infty}d\k
v_n(\k)v_n(\k)=\delta_{mn}.\label{eq:vprop}
\end{eqnarray} Given equations
(\ref{eq:M-tilde},\ref{eq:evectors},\ref{eq:evalues}) we deduce,
\begin{equation}\tilde{M}^{AB}_{mn}v_n(\k)=\tilde{\mu}^{AB}v_m(\k),\end{equation}
with, \begin{eqnarray} \tilde{\mu}(\k) &= &\frac{1}{2\mathrm{cosh}
\frac{\pi \k}{2}-1}, \nonumber\\ \tilde{\mu}^{12}(\k) &=
&\frac{\mathrm{cosh}\frac{\pi \k}{2}+\mathrm{sinh} \frac{\pi
\k}{2}-1}{2\mathrm{cosh}\frac{\pi \k}{2}-1}, \nonumber\\
\tilde{\mu}^{21}(\k) &= &\frac{\mathrm{cosh}\frac{\pi
\k}{2}-\mathrm{sinh}\frac{\pi \k}{2}-1}{2\mathrm{cosh}\frac{\pi
\k}{2}-1}.\label{eq:tevalues}\end{eqnarray} Apparently, the
eigenvalues of $\tilde{M}$ lie in the range $(0,1]$ and are doubly
degenerate for $\k\neq 0$ since $v_n(\k)$ and $v_n(-\k)$ are
linearly independent but have the same eigenvalue:
$\tilde{\mu}(\k)=\tilde{\mu}(-\k)$. The $\k=0$ eigenvalue
$\tilde{\mu}(0)=1$ is the exception, having only one twist odd
eigenvector $v_{2n-1}(0)$.

We will find it helpful to decompose the set of functions
$v_n(\k)$ into odd/even components, $v_{2n}(\k)$ and
$v_{2n-1}(\k)$. By symmetry, $v_{2n}(0)=0$. The ghost matrices
$\tilde{M}^{AB}$ act on $v_{2n}, v_{2n-1}$ as follows,
\begin{eqnarray}\tilde{M}^{AB}_{2n,2m}v_{2m}(\k)&=&\half[
\tilde{\mu}^{AB}(\k)+\tilde{\mu}^{BA}(\k)]v_{2n}(\k),\nonumber\\
\tilde{M}^{AB}_{2n-1, 2m}v_{2m}(\k)&=&\half[
\tilde{\mu}^{AB}(\k)-\tilde{\mu}^{BA}(\k)]v_{2n-1}(\k),\nonumber\\
\tilde{M}^{AB}_{2n-1,2m-1}v_{2m-1}(\k)&=&\half[
\tilde{\mu}^{AB}(\k)+\tilde{\mu}^{BA}(\k)]v_{2n-1}(\k),\nonumber\\
\tilde{M}^{AB}_{2n, 2m-1}v_{2m-1}(\k)&=&\half[
\tilde{\mu}^{AB}(\k)-\tilde{\mu}^{BA}(\k)]v_{2n}(\k),
\end{eqnarray} Now \EQN{vprop} takes the form,
$$v_{2m-1}(\k)v_{2m-1}(\k')=\half\delta(\k-\k')\ \ \ \ \int_0^\infty
\ d\k\ v_{2m-1}(\k)v_{2n-1}(\k)=\half\delta_{2m-1,2n-1}$$
\begin{equation}v_{2m}(\k)v_{2m}(\k')=\half\delta(\k-\k')\
\ \ \ \int_0^\infty \ d\k\
v_{2m}(\k)v_{2n}(\k)=\half\delta_{2m,2n}\label{eq:vprop2}\end{equation}
We have now gathered all the technology we need to diagonalize the
quadratic form appearing in the reduced three string vertex. We
will write the vertex in terms of the following continuous
collection of oscillators: $$\bo(\k) \equiv
\sqrt{2}\sum_{n=1}^\infty \frac{v_{2n-1}(\k)}{\sqrt{2n-1}}
b_{2n-1}\ \ \ \ \be(\k)\equiv i\sqrt{2} \sum_{n=1}^{\infty}
\frac{v_{2n}(\k)}{\sqrt{2n}}b_{2n},$$
\begin{equation}\co(\k)\equiv
\sqrt{2}\sum_{n=1}^{\infty} \sqrt{2n-1} v_{2n-1}(\k)c_{2n-1}\ \ \
\ \ce(\k)\equiv i\sqrt{2}\sum_{n=1}^\infty
\sqrt{2n}v_{2n}(\k)c_{2n}.\label{eq:modes}\end{equation} These
satisfy a $bc$ oscillator algebra with a continuous mode label:
\begin{equation}\{\be^{+}(\k),\ce(\k')
\}=\{\bo^{+}(\k),\co(\k')
\}=\delta(\k-\k').\label{eq:con-bc}\end{equation} All other
distinct anticommutators between the $\beta$s and $\chi$s vanish.
To avoid cluttered notation, we will often write
$\vec{\beta}(\k)=\pmatrix{\be(\k) \cr \bo(\k)}$, and similarly for
$\vec{\chi}$, and sometimes suppress the ``index'' $\k$ whenever
this won't lead to confusion. We can invert \EQN{modes} using
\EQN{vprop2},
$$b_{2n-1}^+=\sqrt{2}\sqrt{2n-1} \int_0^\infty d\k\ v_{2n-1}\bo^+\ \ \ \
b_{2n}^+=i\sqrt{2}\sqrt{2n} \int_0^\infty d\k\ v_{2n}\be^+,$$
\begin{equation}c_{2n-1}^+=\sqrt{2}\frac{1}{\sqrt{2n-1}} \int_0^\infty
d\k\ v_{2n-1}\co^+\ \ \ \ c_{2n}^+=i\sqrt{2}\frac{1}{\sqrt{2n}}
\int_0^\infty d\k\ v_{2n}\ce^+.\label{eq:invmodes}\end{equation}
Now simply plug these formulas into the quadratic form,

\bigskip \noindent$b^{A+}_m\frac{1}{\sqrt{m}}\tilde{N}^{AB}_{mn}\sqrt{n}c^{B+}_n$
\begin{eqnarray}
&=&
b_{2m}^{A+}\sqrt{\frac{2n}{2m}}\tilde{M}^{AB}_{2m,2n}c_{2n}^{B+}+
b_{2m}^{A+}\sqrt{\frac{2n-1}{2m}}\tilde{M}^{AB}_{2m-1,2n}c_{2n-1}^{B+}\nonumber\\
&\ &\ \
-b_{2m-1}^{A+}\sqrt{\frac{2n-1}{2m}}\tilde{M}^{AB}_{2m,2n-1}c_{2n}^{B+}
-b_{2m-1}^{A+}\sqrt{\frac{2n-1}{2m-1}}\tilde{M}^{AB}_{2m-1,2n-1}c_{2n-1}^{B+}
\nonumber\\
&=&\sum_{m,n=1}^{\infty}\left[-2\int_0^\infty d^2 \k\
\be^{A+}(\k)\ce^{B+}(\k')v_{2m}(\k)\tilde{M}^{AB}_{2m,2n}v_{2n}(\k')\right.
\nonumber\\ &\ &+2i\int_0^\infty d^2 \k\
\be^{A+}(\k)\co^{B+}(\k')v_{2m}(\k)\tilde{M}^{AB}_{2m,2n-1}v_{2n-1}(\k')
\nonumber\\ &\ &\ -2i\int_0^\infty d^2 \k\
\bo^{A+}(\k)\ce^{B+}(\k')v_{2m-1}(\k)\tilde{M}^{AB}_{2m-1,2n}v_{2n}(\k')
\nonumber\\ &\ &\left.-2\int_0^\infty d^2 \k\
\bo^{A+}(\k)\co^{B+}(\k')v_{2m-1}(\k)\tilde{M}^{AB}_{2m-1,2n-1}v_{2n-1}(\k')\right]
\nonumber\\ &=&-\half\int_0^\infty d\k\
\big[(\tilde{\mu}^{AB}+\tilde{\mu}^{BA})\vec{\beta}^{B+}\cdot\vec{\chi}^{B+}
+(\tilde{\mu}^{AB}-\tilde{\mu}^{BA})\vec{\beta}^{A+}
\cdot\sigma_y\cdot\vec{\chi}^{B+}\big].\nonumber
\end{eqnarray} Where $\sigma_y=\pmatrix{0 & -i \cr i &
0}$. Explicitly, then, the reduced three string vertex can be
written in the following form:
\begin{eqnarray}|\hat{V}_3\rangle&=&\mathrm{exp}\left[\half\int_0^\infty
d\k\
\left\{2\tilde{\mu}(\vec{\beta}^{1+}\cdot\vec{\chi}^{1+}+\mathrm{cyclic.})
\right.\right.\nonumber\\ &\ &+(\tilde{\mu}^{12}+\tilde{\mu}^{21})
(\vec{\beta}^{1+}\cdot\vec{\chi}^{2+}+\vec{\beta}^{2+}\cdot\vec{\chi}^{1+}
+\mathrm{cyclic.})\nonumber\\ &\
&\left.\left.+(\tilde{\mu}^{12}-\tilde{\mu}^{21})
(\vec{\beta}^{1+}\cdot\sigma_y\cdot\vec{\chi}^{2+}-
\vec{\beta}^{2+}\cdot\sigma_y\cdot\vec{\chi}^{1+}+\mathrm{cyclic})\right\}\right]
\nonumber\\ &\ &\ \ \ \ \times|0\rangle_1|0\rangle_2|0\rangle_3
\label{eq:diagonalvertex}\end{eqnarray} So, we have succeeded
diagonalizing the quadratic form in the vertex. This form makes it
clear that the reduced star algebra in the fermionic ghost sector
is a continuous tensor product\footnote{When we talk about taking
a ``tensor product'' of two algebras in this paper, we mean it in
a special sense. The naive multiplication rule for the tensor
product of two algebras is $(a\otimes c)(b\otimes d)= ab\otimes
cd$ for $a,\ b$ in the first algebra and $c,\ d$ in the second.
This is not the natural definition here, since we're dealing with
Grassmann quantities. Rather, we should define $(a\otimes
c)(b\otimes d)= (-1)^{G(b)+G(c)}ab\otimes cd$, where $G(b)$ and
$G(c)$ are the Grassmann rank of $b$ and $c$, respectively. With
this definition, note in particular that $Cl_{m,n}\otimes Cl_{p,q}
= Cl_{m+p,n+q}$.}, from $\k=0$ to $\infty$, of mutually commuting
algebras. Our task is now to understand the nature of these
algebras.

\section*{IV. Moyal structure in the ghost vertex}
On general grounds we expect the star product in the fermionic
ghost sector to be closely related to a Moyal product. Somehow we
need to translate the Fock space representation of the star
product, \EQN{diagonalvertex}, into an integral kernel
representation so we can explicitly compare this to the Moyal
product, \EQN{kernel}. The crucial step is to construct
``position'' and ``momentum'' operators out of the $bc$ oscillator
algebra, and use the eigenvectors of these operators to construct
wavefunctions out of the Hilbert space states. Performing a
contraction between two states and the three vertex should then
correspond, in a simple way, to a Moyal product between the
associated wavefunctions. So, given a $bc$ oscillator algebra we
define the hermitian operators, \begin{eqnarray}&\
&\hat{x}=\frac{i}{\sqrt{2}}(b-b^+)\ \ \ \ \ \ \ \ \ \
\hat{y}=\frac{1}{\sqrt{2}}(c+c^+) \nonumber\\ &\
&\hat{p}=\frac{i}{\sqrt{2}}(c-c^+)\ \ \ \ \ \ \ \ \ \
\hat{q}=\frac{1}{\sqrt{2}}(b+b^+)\label{eq:xypq}\end{eqnarray} It
is simple to see that these operators are hermitian and satisfy
$\{\hat{x},\hat{p}\}=\{\hat{y},\hat{q}\}=1$, with all other
anticommutators vanishing. As the notation suggests, we will think
of $\hat{x},\ \hat{y}$ as coordinates and $\hat{p},\ \hat{q}$ as
momenta. This is largely a matter of convention, but \EQN{xypq}
was chosen so that Fourier modes of the ghost fields $b(\sigma)$
and $c(\sigma)$ (defined in the next section) would be
coordinates. We can write the eigenstates of $\hat{x},\ \hat{y}$,
\begin{equation}|x,y\rangle=\mathrm{exp}\big[-b^+c^+
+i\sqrt{2} xc^+ +\sqrt{2}b^+y
-ixy\big]|0\rangle.\label{eq:eigenstate}
\end{equation} The inner product of two such states
is, $$\langle x,y|x',y'
\rangle=-2i(x-x')(y-y')=\delta(x'-x)\delta(y'-y).$$ The factor of
$i$ is present to ensure that the inner product is real. The
eigenstates are also complete, \begin{equation}\int dX\
|x,y\rangle\langle x,y|=1,\ \ \ \ dX=-\fraction{i}{2}\
dxdy.\label{eq:measure}\end{equation} Therefore, we may express
the state $|\psi\rangle$ as a wavefunction $\psi(x,y)=\langle
x,y|\psi\rangle$. We put a factor of $\half$ in the measure rather
than $\fraction{1}{\sqrt{2}}$ in the normalization of
$|x,y\rangle$ so that we wouldn't have to keep track of divergent
normalizations when taking infinite tensor products of such
states, as we will need to do in the next section. Given a
3-vertex $|V\rangle$, allowing us to define a product between
states, $|\psi*\phi\rangle=\langle\psi| \langle\phi|V\rangle$, it
follows that we can write $\psi*\phi(X)=\int dX^1 dX^2\
K(X^1,X^2,X)\psi(X^1)\phi(X^2)$ with $K(X^1, X^2,X)=\langle
X|\langle X^1|\langle X^2||V\rangle$, denoting $X=(x,y)$.

Taking a hint from \EQN{con-bc}, let us consider the tensor
product of two $bc$ oscillator algebras, which we will write for
illustrative purposes $(\bev,\cev)$ and $(\bod,\cod)$. A kernel of
particular interest is, $$K(X^1,X^2,X^3)=16Ng^4\mathrm{exp}\left[
-i \ \vec{x}^A \cdot W^{AB}\cdot \vec{y}^B\right], $$ with,
\begin{equation}W^{AB}=g^{-1}\pmatrix{0 & -\sigma_y &
\sigma_y\cr \sigma_y & 0 & -\sigma_y \cr -\sigma_y & \sigma_y & 0
}^{AB},\label{eq:star-kernel}\end{equation} The index $A$ ranges
from 1 to 3 and $\vec{x}=\pmatrix{\xe \cr \xo}$ and likewise for
$\vec{y}$. We can also rewrite the quadratic form as, $$-i \
\vec{x}^A \cdot W^{AB}\cdot \vec{y}^B = -\xo^A K^{AB} \ye^B +\xe^A
K^{AB} \yo^B $$ where $K^{AB}$ is the matrix in \EQN{kernel}
defining the kernel for the Moyal product. Therefore,
\EQN{star-kernel} is just a kernel for the tensor product of two
Clifford Algebras, up to some normalization\footnote{According to
\EQN{kernel} the conventional normalization would be $N=1$. The
factor of $16$ in front of the above kernel is necessary to cancel
four factors of $\half{}$ which go into our definition of the
measure \EQN{measure}.}, with $(\xe,\yo)$ and $(\xo,\ye)$ forming
canonical Moyal pairs.

Let us derive the 3-vertex corresponding to the kernel
\EQN{star-kernel}. Assume it is of the form, $$|V\rangle =
\mathrm{exp}[\vec{b}^{A+}\cdot V^{AB}\cdot
\vec{c}^{B+})]|0\rangle_1|0\rangle_2|0\rangle_3$$ Then we have,
\begin{eqnarray}K(X^1,X^2,X^3)&=&\langle 0|^{3}
\mathrm{exp}\big[\vec{b}^A\cdot\vec{c}^A +i\sqrt{2} \vec{x}^A\cdot
\vec{c}^A -\sqrt{2}\vec{b}^A \cdot \vec{y}^A -i\vec{x}^A \cdot
\vec{y}^A\big]\nonumber\\ &\ &\ \ \ \ \ \ \times
\mathrm{exp}[\vec{b}^{A+} \cdot V^{AB} \cdot
\vec{c}^{B+})]|0\rangle^3,\nonumber\\
&=&\mathrm{det}(1+V)\mathrm{exp}\left[-i\vec{x}^A \cdot
\left(\frac{1-V}{1+V}\right)^{AB}\cdot \vec{y}^B\right].\nonumber
\end{eqnarray}
This implies,
\begin{equation}N=\frac{\mathrm{det}(1+V)}{16g^4}=\left(\frac{2}{g^2+3}\right)^2,\
\ \ \ \
V^{AB}=\left(\frac{1-W}{1+W}\right)^{AB}.\label{eq:NW}\end{equation}
Plugging in \EQN{star-kernel} and working through some matrix
algebra we find:
\begin{equation}V^{AB}=\frac{1}{g^2+3}\pmatrix{g^2-1 &
2+2g\sigma_y & 2-2g\sigma_y\cr 2-2g\sigma_y & g^2-1 & 2+2g\sigma_y
\cr 2+2g\sigma_y & 2-2g\sigma_y & g^2-1}^{AB}
\label{eq:V}\end{equation} The vertex is then becomes,
\begin{eqnarray}|V\rangle&=&
\mathrm{exp}\left[\frac{g^2-1}{g^2+3}(\vec{b}^1 \cdot\vec{c}^1 +
\mathrm{cyclic.})+\frac{2}{g^2+3}(\vec{b}^1
\cdot\vec{c}^2+\vec{b}^2 \cdot\vec{c}^1
+\mathrm{cyclic})\right.\nonumber\\ &\ &\ \ \ \ \
\left.+\frac{2g}{g^2+3}( \vec{b}^1 \cdot\sigma_y \cdot \vec{c}^2
-\vec{b}^2 \cdot\sigma_y \cdot \vec{c}^1+
\mathrm{cyclic})\right]|0\rangle_1 |0\rangle_2
|0\rangle_3\label{eq:modelvertex}\end{eqnarray} A quick comparison
with \EQN{diagonalvertex} shows that this vertex is in the same
form as the diagonalized three string vertex. However, if the two
vertices are really the same, we must be able to find a metric
$g(\k)$ satisfying,
$$\tilde{\mu}(\k)=\frac{g(\k)^2-1}{g(\k)^2+3}, \ \ \ \ \
\half(\tilde{\mu}^{12}(\k)+\tilde{\mu}^{21}(\k))=\frac{2}{g(\k)^2+3},$$
$$
\half(\tilde{\mu}^{12}(\k)-\tilde{\mu}^{21}(\k))=\frac{2g(\k)}{g(\k)^2+3}.$$
Remarkably, a solution exists:
\begin{equation}g(\k)=\mathrm{coth}\fraction{\pi
\k}{4}.\label{eq:metric}\end{equation} Therefore, the reduced star
product in the fermionic ghost sector is given by a continuous
tensor product---from $\k>0$ to $\infty$---of a {\it pair} of
Clifford Algebras $\C\otimes Cl_{1,1}$, the first having a metric
$+g(\k)$ and the second $-g(\k)$. Next section we will explicitly
construct a path integral representation for the reduced star
product using a continuous product of kernels of the form
\EQN{star-kernel}.

It is interesting to compare \EQN{metric} to the analogous result
in the matter sector, where in ref.\cite{Moyal} they found a
noncommutativity parameter $\theta(\k)$ scaling as
$\tanh(\fraction{\pi \k}{4})$. The fact that $\theta(\k)$ vanishes
at $\k=0$ is significant, indicating the presence of a commutative
coordinate. Here we have no purely anticommutative coordinate, but
instead an infinitely nonanticommutative algebra at $\k=0$. This
seems most peculiar, and its relevance is not immediately obvious.

We should mention that our definition of the wavefunction
$\psi(x,y)=\langle x,y|\psi\rangle$ is crucial for establishing
the above result. If instead we had defined $\psi$ to depend on
the eigenvalues of $\hat{p}$ or $\hat{q}$, we would find that the
vertex describing the Moyal product of these functions would not
be of the form \EQN{diagonalvertex}. This is hardly a surprise,
since the reduced star product cannot be a Moyal product in
position and momentum space simultaneously. Still, one wonders why
the Moyal structure emerges in ``position'' space, especially
considering that our definition of position versus momentum seemed
somewhat arbitrary. The reason comes down to our choice of
multiplicative phase in the definition of $\beta,\ \chi$ in
\EQN{modes}. If we redefined the phase so that $\beta\to-i\beta$
and $\chi\to i\chi$ then the three string vertex would actually
correspond to fermionic Moyal star multiplication in ``momentum
space.'' Clearly there is no physical content in our choice of
conventions.

\section*{V. Path integral kernel representation of
the reduced star product} In the last section we argued that the
diagonalized vertex \EQN{diagonalvertex} actually represents a
continuous tensor product of Clifford Algebras, $\C\otimes
Cl_{2,2}$, up to some normalization (except at $\k=0$). In this
section we would like to translate the reduced star product from
an oscillator vertex into a path integral kernel, analogous to
\EQN{kernel}. Let us begin by defining position and momentum modes
as in \EQN{xypq},
\begin{eqnarray}&\
&\hat{x}_n=\frac{i}{\sqrt{2}}(b_n-b_n^+)\ \ \ \ \ \ \ \ \ \
\hat{y}_n=\frac{1}{\sqrt{2}}(c_n+c_n^+) \nonumber\\ &\
&\hat{p}_n=\frac{i}{\sqrt{2}}(c_n-c_n^+)\ \ \ \ \ \ \ \ \ \
\hat{q}_n=\frac{1}{\sqrt{2}}(b_n+b_n^+)\label{eq:x_n}\end{eqnarray}
and the associated eigenstate,
\begin{equation}|x_n,
y_n\rangle=\mathrm{exp}\big[-b^+_n c^+_n +i\sqrt{2} x_n c^+_n
+\sqrt{2}b^+_ny_n -ix_n y_n\big]|0\rangle.\label{eq:x_n-eig}
\end{equation} ($|0\rangle$ again is the vacuum
annihilated by $b_n,\ c_n,\ n>0$.) The operators $x_n$ and $y_n$
are the Fourier modes of the ghost fields,
\begin{eqnarray}&\ &
b(\sigma)=i\sqrt{2}\sum_{n=1}^{\infty} \hat{x}_n \sin
n\sigma\nonumber\\
&\ &c(\sigma)=c_0+\sqrt{2}\sum_{n=1}^{\infty} \hat{y}_n \cos
n\sigma,\label{eq:bcfields}
\end{eqnarray} where,
$$c^{\pm}(\sigma)=c(\sigma)\pm i\pi_b(\sigma)\ \ \ \
b_{\pm\pm}(\sigma)=\pi_c(\sigma)\mp i b(\sigma)$$ The string field
in the fermionic ghost sector is often written as a functional of
$b(\sigma)$ and $c(\sigma)$,
\begin{equation}\Psi[b(\sigma),
c(\sigma)]=\Psi[c_0, x_n, y_n]= \langle c_0| \otimes\langle x_n,
y_n||\Psi\rangle \label{eq:string-functional}\end{equation} where
$|c_0\rangle=\exp[\hat{b}_0 c_0]|+\rangle$ is the eigenstate of
$\hat{c}_0$. Remember that we are working in Siegel gauge, which
means,
$$b_0|\Psi\rangle=0\ \ \Rightarrow\ \der{c_0}\Psi[c_0,
x_n, y_n]=0\ \ \Rightarrow\ \Psi=\Psi[x_n, y_n].$$ As implied by
our discussion in the last section, the Moyal structure of the
reduced star product is clearly manifest when the string field is
expressed as a functional of the eigenvalues of the operators,
\begin{eqnarray}&\
&\hat{x}_\e(\k)=\fraction{i}{\sqrt{2}}(\be(\k)-\be^+(\k))=
-\sqrt{2}\sum_{n=1}^{\infty} \frac{1}{\sqrt{2n}}v_{2n}(\k)\hat{q}_{2n}\nonumber\\
&\ &\hat{x}_\o(\k)=\fraction{i}{\sqrt{2}}(\bo(\k)-\bo^+(\k))=
\sqrt{2}\sum_{n=1}^\infty \frac{1}{\sqrt{2n-1}}v_{2n-1}(\k)\hat{x}_{2n-1}\nonumber\\
&\ &\hat{y}_\e(\k)=\fraction{1}{\sqrt{2}}(\ce(\k)+\ce^+(\k))=
\sqrt{2}\sum_{n=1}^\infty \sqrt{2n} v_{2n}(\k)\hat{p}_{2n}\nonumber\\
&\ &\hat{y}_\o(\k)=\fraction{1}{\sqrt{2}}(\co(\k)+\co^+(\k))=
\sqrt{2}\sum_{n=1}^\infty
\sqrt{2n-1}v_{2n-1}(\k)\hat{y}_{2n-1}\label{eq:cont-xy}
\end{eqnarray}
The string field as it is usually written in
\EQN{string-functional} is not a functional of the eigenvalues of
\EQN{cont-xy}. We need to perform a Fourier transform on the even
coordinates:
\begin{equation}\tilde{\Psi}[x_{2n-1}, y_{2n-1}, p_{2n},
q_{2n}]=\int
\left(\prod_{m=1}^{\infty}\fraction{1}{i}dx_{2m}dy_{2m}
e^{p_{2n}x_{2n}+q_{2n} y_{2n} }\right) \Psi[x_n, y_n]
.\label{eq:half-Fourier}\end{equation} Now define,
$$\Psi^M[\vec{x}(\k), \vec{y}(\k)]=\tilde{\Psi}[x_{2n-1}, y_{2n-1}, p_{2n},
q_{2n}]$$ This is the form of the string field we require.

From \EQN{x_n-eig} it follows that the ground state of the ghost
Fock space is the functional,
\begin{eqnarray}\Psi_{|0\rangle}[x_n,y_n]&=&e^{-i x_ny_n}\
\nonumber\\ \tilde{\Psi}_{|0\rangle}[x_{2n-1},y_{2n-1},
p_{2n},q_{2n}]&=&\exp\left[-i(
x_{2n-1}y_{2n-1}+p_{2n}q_{2n})\right]
\label{eq:ground-state}\end{eqnarray} The states $|x_n,y_n\rangle$
are normalized, just as in \EQN{measure}, so that they furnish a
resolution of the identity with respect to a measure given as an
infinite product of $-\fraction{i}{2}dx_n dy_n$. Consistently, we
have,
\begin{eqnarray}&\ &\langle
0|0 \rangle = 1 = \int \left(\prod_{n=1}^{\infty}\fraction{1}{2i}
dx_n dy_n\right)e^{-2ix_ny_n}\nonumber\\ &\ &\ \ \ \ \ \ \ \ \ \ \
\ =\int \prod_{n=1}^{\infty}\fraction{1}{2i} dx_{2n-1}
dy_{2n-1}\fraction{1}{2i}dp_{2n}dq_{2n}e^{-2i x_{2n-1} y_{2n-1}-2i
p_{2n} q_{2n}}.\nonumber \end{eqnarray} From \EQN{cont-xy} we see
that the ground state can also be expressed as a functional of
$\vec{x}(\k), \vec{y}(\k)$ (again we will suppress the $\k$ index
when this is unambiguous),
\begin{equation}\Psi^M_{|0\rangle}[\vec{x}(\k),
\vec{y}(\k)] =\exp\left[-i\int_0^\infty d\k(\xo
\yo-\xe\ye)\right]\label{eq:ground} .\end{equation} Let us define
a path integral measure over $\vec{x}$ and $\vec{y}$, so that this
ground state functional has unit norm:
\begin{equation}\int
[dx dy]\exp\left(-2i\int_0^\infty d\k[x y+ j_x y +x j_y]
\right)\nonumber=\exp\left(2i\int_0^\infty d\k\ j_x j_y
\right),\label{eq:path-integral}\end{equation} from which we can
see that,
$$\int[d\vec{x} d\vec{y}]\left|\Psi^M_{|0\rangle}[\vec{x},
\vec{y}]\right|^2=1$$ We will also need to consider the path
integral, \begin{equation}\int [dx dy]\exp\left(-2i\int_0^\infty
d\k\ G x y \right)=\exp\left(\delta(0)\int_0^\infty d\k\
\ln[G]\right)\label{eq:path-integral2}\end{equation} The infinite
$\delta(0)$ is the same as the divergent factor $\frac{\ln
L}{2\pi}$, with $L$ is the ``level regulator,'' discussed in
\cite{Spectroscopy,Okuyama,Moyal}. We can see intuitively that
this divergent factor is necessary, since rescaling a particular
$x(\k')$ by $G$ should scale the whole path integral by $G$. This
is consistent with setting
$G(\k)=1+(G-1)\delta(\k-\k')/\delta(0)$, and plugging into
\EQN{path-integral2}.

From \EQN{x_n-eig} we can derive the eigenstate of the operators
$\hat{x}_\e,\ \hat{x}_\o,\ \hat{y}_\e$, and $\hat{y}_\o$ through
taking Fourier transforms and the appropriate linear combinations:
\begin{equation}|\vec{x},
\vec{y}\rangle =\mathrm{exp}\left[\int_0^\infty\left(
-\vec{\beta}^+ \cdot\vec{\chi}^+ + i\sqrt{2}
\vec{x}\cdot\vec{\chi}^+ + \sqrt{2}\vec{\beta}^+ \cdot\vec{y}
-i\vec{x}\cdot\vec{y}\right)\right]|0\rangle.\label{eq:x(k)-eig}
\end{equation} We can then calculate the wavefunctional by performing
the contraction, $$\Psi^M[\vec{x}, \vec{y}]=\langle c_0|\otimes
\langle \vec{x}, \vec{y}||\Psi\rangle $$ This is consistent with
\EQN{ground}. Therefore, with the choice of measure
\EQN{path-integral}, we have a resolution of the identity,
\begin{equation} \int [d\vec{x}\ d\vec{y}]\
|\vec{x}, \vec{y}\rangle \langle\vec{x}, \vec{y}| =1
\end{equation} Armed with this we can easily translate
the three string vertex \EQN{diagonalvertex} into the kernel of
the path integral representing the reduced star product
($\vec{X}=(\vec{x},\vec{y})$):
\begin{equation}\Phi^M*_{b_0}\Psi^M[\vec{X}^3]=\int\prod_{A=1,2}
[d\vec{x}^A\ d\vec{y}^A] \Phi^{M*}[\vec{X}^1]
\Psi^{M*}[\vec{X}^2]K[\vec{X}^1, \vec{X}^2,\vec{X}^3]
\label{eq:path-int-kernel}
\end{equation} where $$K[\vec{X}^1,
\vec{X}^2,\vec{X}^3]=\langle \vec{X}^1|\langle \vec{X}^2| \langle
\vec{X}^3||\hat{V}_3\rangle $$ Performing the contraction, we
get,\begin{equation}K[\vec{X}^1, \vec{X}^2,\vec{X}^3]
=\fraction{1}{8}\exp\left[-i\int_0^\infty d\k\vec{x}^A\cdot
W^{AB}\cdot \vec{y}^B -2\delta(0)\int_0^\infty d\k\ \ln
\left(\frac{3+g^2}{8g^2}\right)
\right]\label{eq:full-kernel}\end{equation} where $W^{AB}(\k)$ the
same matrix as in \EQN{star-kernel} with
$g=g(\k)=\coth\fraction{\pi \k}{4}$. This is exactly as
anticipated in the previous section. Note that the normalization
factor relating the reduced star product to a canonically
normalized fermionic Moyal star is,
$$\mathcal{N}=\fraction{1}{8}\exp\left[-2\delta(0)\int_0^\infty
d\k\ \ln \fraction{1}{2}\left(3+g^2\right) \right].$$
$\mathcal{N}$ is the analogy of the constant $N$ calculated in
\EQN{NW}. In ref.\cite{Moyal} it was suggested that $\mathcal{N}$
might cancel a corresponding divergent normalization in the matter
sector kernel. However, explicitly calculating the product of the
two factors,
$$\mathcal{N}C'=8\exp\left[\delta(0)\left\{D\int_0^\infty
d\k\ \ln \fraction{1}{8}\left(12+\theta^2\right)-2\int_0^\infty
d\k\ \ln \fraction{1}{2}\left(3+g^2\right)\right\} \right],$$
where $\theta(\k)=2\tanh\frac{\pi \k}{4}$ is the bosonic
noncommutativity parameter\cite{Moyal}, we see that the product is
infinite, at least for $D\neq 2$, though the ghost contribution
does temper the divergence as expected. This implies that the
reduced star algebra (with the matter part) does not really define
an algebra of bounded operators; the reduced star product of two
string fields of finite norm will generally yield a state of
infinite norm. Of course, the fact that the reduced star algebra
is unbounded does not necessarily imply that the {\it full} star
algebra is unbounded\footnote{Thanks to L. Rastelli for pointing
this out to me.}; it might be possible that the full star algebra
is bounded in $D=26$. This is an important question which we leave
for future inquiry.

Let us summarize our results as follows: the reduced star product
can be written,
$$\Phi^M*_{b_0}\Psi^M[\vec{x}(\k),\vec{y}(\k)]=\mathcal{N}
\Phi^M\star\Psi^M[\vec{x}(\k),\vec{y}(\k)],$$ with $\star$ the
canonically normalized Moyal product satisfying,
\begin{equation}
\{\xe(\k),\yo(\k')\}_\star = -\{\xo(\k),\ye(\k')\}_\star =
2\coth\fraction{\pi \k}{4}\delta(\k-\k')\ \
\label{eq:cont-Moyal}\end{equation} All other anticommutators
between $\vec{x}(\k)$ and $\vec{y}(\k)$ vanish. This completes the
Moyal formulation of the reduced star product in the fermionic
ghost sector.

\section*{VI. Kinetic and ghost number operators}
In this section we will formulate some important operators in
string field theory using the continuous oscillator basis
introduced in \EQN{modes}. Sofar we have established that the
reduced star algebra can be formulated as an infinite tensor
product of Clifford Algebras, but eventually we are interested in
the {\it full} star product, which as explained before is related
to the reduced star product in Siegel gauge through the action of
the operator,
\begin{equation}\mathcal{Q}=c(\fraction{\pi}{2})=
c_0-(c_2+c_2^+)+(c_4+c_4^+)-...,\label{eq:Q}\end{equation} which
also happens to be the canonical choice of pure ghost kinetic
operator in VSFT (up to a possibly infinite
normalization)\cite{Hata,Sen,Okuyama2}. Clearly we would rather
not translate back into the old oscillator basis whenever we need
to calculate the full star product from the reduced one, so it is
advantageous to formulate $\mathcal{Q}$ within the language
developed here. Write,
$$\mathcal{Q}=c_0-i\sqrt{2}\int_0^\infty d\k K(\k)(\ce(\k)-\ce^+(\k))$$
where,
$$K(\k)=\sum_{n=1}^{\infty}(-1)^n
\frac{v_{2n}(\k)}{\sqrt{2n}}= \frac{2-
e^{-\k\tan^{-1}i}-e^{-\k\tan^{-1}(-i)}}{2\sqrt{2\k\sinh\fraction{\pi
\k}{2}}}$$ We took the liberty of evaluating the sum using
\EQN{genr-f}. Unfortunately this expression is undefined, since
the inverse tangent has simple poles at $\pm i$. We can regulate
it by replacing $\tan^{-1}i$ by $\tan^{-1}ia$ and taking the limit
$a\to 1$:
\begin{eqnarray}K(\k)=\lim_{a\to 1} \frac{2-
e^{-i\k\tanh^{-1}a}-e^{i\k\tanh^{-1}a}}{ 2
\sqrt{2\k\sinh\fraction{\pi \k}{2}}}\nonumber =
\frac{2}{\sqrt{2\k\sinh\fraction{\pi \k}{2}}}\lim_{w\to \infty}
\sin^2(w \k) \label{eq:divergence}\end{eqnarray} but this doesn't
help since the above expression oscillates with infinite frequency
and does not converge in the limit. However, we should really
think of $K(\k)$ as an object which properly belongs under an
integral. Consider, for instance,
\begin{eqnarray}\int_a^b d\k\ \k^{2n}
\sin^2(w\k)&=&\half\int_a^b d\k\
\k^{2n}-\half\left(-\fraction{1}{4}\right)^n\frac{\partial^{2n}}{\partial
w^{2n}} \int_a^b d\k\ \cos(2w\k)\nonumber\\ &=& \half\int_a^b d\k\
\k^{2n}-\half\left(-\fraction{1}{4}\right)^n\frac{\partial^{2n}}{\partial
w^{2n}} \frac{\cos(2wb)-\cos(2wa)}{2w}.\nonumber \\
\nonumber
\end{eqnarray} Notice that the second term on the
right hand side, for all $n\geq 0$, dies off for large $w$ at
least as $1/w$. A similar argument follows for odd powers of $\k$.
Therefore, $$\lim_{w\to\infty}\int_a^b d\k\
f(\k)\sin^2(w\k)=\half\int_a^b d\k\ f(\k)$$ for any function $f$
with a local Taylor expansion. Taking this into account, we can
write the following expression for $\mathcal{Q}$:
\begin{equation}\mathcal{Q}=c_0-\sqrt{2}\int_0^\infty\frac{d\k\
\hat{p}_\e(\k)}{\sqrt{\k\sinh\fraction{\pi
\k}{2}}}\label{eq:cont-Q}\end{equation} where
$\hat{p}_\e=\fraction{i}{\sqrt{2}}(\ce-\ce^+)$ \footnote{The
apparent pole in the integrand near $\k=0$ is illusory since
$\hat{p}^\e(\k)$ vanishes linearly near $\k=0$.}. In the spirit of
noncommutative field theory, one might like to express the action
of $\mathcal{Q}$ on a state as a star-(anti)commutator with the
associated wavefunctional. We can do this easily by writing the
state as a functional of $\vec{x}(\k),\vec{y}(\k)$, in which case
we represent the action of $\hat{p}^\e(\k)$ as the functional
derivative $\delta/\delta \xe(\k)$. Define the funtional/operator,
\begin{equation}\mathcal{Q}^M=-\half \int_0^\infty d\k\
\left(\frac{\sinh\fraction{\pi \k}{4}}{\k\cosh^3\fraction{\pi
\k}{4}}\right)^{\frac{1}{2}}\yo(\k).\label{Q-M}\end{equation} This
allows us to write, invoking \EQN{derivative-commutator} and
\EQN{cont-Moyal},
\begin{equation}\mathcal{Q}\big(\Psi^M\big)=c_0\Psi^M +\mathcal{Q}^M\star
\Psi^M + (-1)^{G(\Psi)} \Psi^M \star \mathcal{Q}^M.
\label{eq:Q-state}\end{equation} where $\Psi^M=
\Psi^M[\vec{x}(\k),\vec{y}(\k)]$ and $G(\Psi)$ denotes the ghost
number\footnote{Note that the ghost number of the string field is
$1$ plus its Grassmann rank, so \EQN{Q-state} is consistent with
\EQN{cont-Moyal}. This follows since the ground state functional
\EQN{ground-state}, corresponding to the $|-\rangle$ vacuum, is
Grassmann even but has ghost number one.} of $\Psi^M$. Similar
looking expressions for the action of pure ghost kinetic operators
appear in the split string formalism\cite{Gross-Taylor-2}. The
continuous oscillator approach however has a marked advantage over
the split string formalism, since in the latter it seems
impossible to formulate $\mathcal{Q}$ in a well-defined way. This
is both because $\mathcal{Q}$ acts only at one point on the
string, and because this point happens to be the sting midpoint.
In particular, $\mathcal{Q}$ creates a kink in the bosonized ghost
field at $\sigma=\fraction{\pi}{2}$, while $\mathcal{Q}^2$ creates
two kinks; so $\mathcal{Q}$ is apparently not nilpotent.
Furthermore, a configuration with a kink at the midpoint cannot be
encoded with ordinary split string variables $\bar{\phi},\
l(\sigma),\ r(\sigma)$ with
$l(\fraction{\pi}{2})=r(\fraction{\pi}{2})=0$. Though one may
attempt to deal with these problems by regulating and adding more
degrees of freedom at the midpoint, the structure of the split
string formalism, which necessarily treats the midpoint in a
somewhat naive fashion, begins to fall apart. This is in large
part the reason why no satisfactory formulation of VSFT has been
constructed using bosonized ghosts.

An interesting and nontrivial check on the consistency of our
approach and the correctness of our expression for $\mathcal{Q}$
is to verify that $\{\mathcal{Q},\pi_c(\sigma)\}=0$ for
$\sigma\neq \fraction{\pi}{2}$. Note,
\begin{eqnarray}\pi_c(\sigma)&=&\half\big[b_{++}(\sigma)+b_{--}(\sigma)\big]=
b_0+\sqrt{2} \sum_{n=1}^\infty \hat{q}_n\cos
n\sigma\nonumber\\
&=&b_0+2\int_0^\infty d\k\
K_\o(\sigma,\k)\hat{q}_\o(\k)+2\int_0^\infty d\k\
K_\e(\sigma,\k)\hat{x}_\e(\k)\nonumber \end{eqnarray} where,
\begin{eqnarray}K_\e(\sigma,\k)&=&\sum_{n=1}^\infty\sqrt{2n}
\ v_{2n}(\k)\cos 2n\sigma\nonumber\\
&=&\frac{\k}{4\sqrt{2\k\sinh\fraction{\pi \k}{2}}}\left[\frac{\phi
(e^{\k\tan^{-1}\phi}-e^{-\k\tan^{-1}\phi})}{1+\phi^2}+\mathrm{c.c.}
\right]\nonumber\end{eqnarray} where $\phi=e^{i\sigma}$ and c.c.
denotes the complex conjugate of the first term. This equation is
obtained straightforwardly by using the generating function for
the $v_n$s to evaluate the sum. A similar expression defines
$K_\o(\sigma,\k)$ but we will not need it here. Calculate the
anticommutator,
$$\{\mathcal{Q},\pi_c(\sigma)\}=\{c_0,b_0\}-2\sqrt{2}\int_0^\infty
d\k d\k'\
\frac{K^\e(\sigma,\k')\{\hat{p}_\e(\k),\hat{x}_\e(\k')\}}{\sqrt{\k
\sinh\fraction{\pi \k}{4}}}$$ If this vanishes, we must have,
\begin{equation}\int_0^\infty d\k\
\frac{K_\e(\sigma,\k)}{\sqrt{2\k \sinh\fraction{\pi
\k}{2}}}=\frac{1}{4}\label{eq:integral}\end{equation} plugging in
$K_\e$ we are lead to consider the integral,
$$E(\sigma)=\frac{\phi}{4(1+\phi^2)}\int_{-\infty}^\infty
d\k\ \frac{e^{\k\tan^{-1}\phi}}{2\sinh\fraction{\pi \k}{2}}$$ This
expression plus its complex conjugate is the quantity in
\EQN{integral}. Consider,
$$\tan^{-1}\phi=\half\tan^{-1}\left[\frac{\cos\sigma}{1-\sin\sigma}
\right]-\half
\tan^{-1}\left[\frac{\cos\sigma}{1+\sin\sigma}\right]-\frac{i}{4}\ln
\left[\frac{(1-\sin\sigma)^2+\cos^2\sigma}{(1+\sin\sigma)^2+\cos^2\sigma}\right]
$$ The real part of this is always between
$-\fraction{\pi}{2}$ and $\fraction{\pi}{2}$ so $E(\sigma)$ for
$\sigma\neq\frac{\pi}{2}$ is always rendered convergent by the
$\sinh\fraction{\pi \k}{2}$ in the denominator. The imaginary part
is always positive imaginary so we close the contour in the upper
half plane:
\begin{eqnarray}E(\sigma)&=&\frac{2\pi i
\phi}{4(1+\phi^2)}\big[\half \mathrm{Res}(0)+\sum_{n=1}^\infty
\mathrm{Res}(2in)\big]\nonumber\\
&=&\frac{2\pi i \phi}{4(1+\phi^2)}\left[
\frac{1}{2\pi}+\sum_{n=1}^\infty
\frac{(-1)^n}{\pi}\left(\frac{1+i\phi}{1-i\phi}\right)^n\right]
\nonumber\\ &=&\frac{\phi^2}{4(1+\phi^2)}\nonumber
\end{eqnarray}
Above we took the principle value of the integral, though actually
the final result is the same regardless of our contour
prescription. So we are left to check,
$$E(\sigma)+\bar{E}(\sigma)=\frac{\phi^2}{4(1+\phi^2)}+\frac{\bar{\phi}^2}{4(1+\bar{\phi}^2)}
=\frac{\phi^2}{4(1+\phi^2)}+\frac{1}{4(1+\phi^2)}=\frac{1}{4}$$
just as required. This argument does not work (fortunately) at
$\sigma=\frac{\pi}{2}$, since $E(\sigma)$ is divergent there.
Obviously, we expect a delta function at $\frac{\pi}{2}$.

Another interesting operator we should consider is the ghost
number operator, \begin{eqnarray}&\
&G=\fraction{3}{2}+\half(c_0b_0-b_0c_0)+
\sum_{n=1}^{\infty}(c_n^+ b_n-b_n^+ c_n)\nonumber \\
&\ &\ \ \ =1+\int_0^\infty
d\k(\vec{\chi}^+\cdot\vec{\beta}-\vec{\beta}^+\cdot\vec{\chi})
\label{ghost-number}\end{eqnarray} In the second expression we
assumed Siegel gauge. One of the crucial properties of Witten's
star product is that the ghost number of the product of two string
fields is the sum of the ghost numbers of the fields individually;
this is what makes the star product of fields similar to the wedge
product of differential forms. Let us see how this property
emerges in the present formalism. Note that the reduced star
product must be additive in $G_0=G-1$:
\begin{eqnarray}G_0(A*_{b_0}B)&=&G(A*_{b_0}B)-1
=G(A*B)-2 =G(A)+G(B)-2\nonumber\\
&=&G_0(A)+G_0(B)\nonumber
\end{eqnarray} which we
call the reduced ghost number. Acting on a string field expressed
as a functional of $\vec{x}(\k),\vec{y}(\k)$, the reduced ghost
number operator is, $$G_0=\int_0^\infty
d\k\left(\vec{y}(\k)\cdot\frac{\delta}{\delta \vec{y}(\k)}-
\vec{x}(\k)\cdot\frac{\delta}{\delta \vec{x}(\k)}\right)$$ It
turns out that $G_0$ acts as a derivation of the $\star$ algebra.
We can argue this as follows. The kernel derived in \EQN{kernel}
is easily seen to satisfy, $$\sum_{A=1}^3
\left(y^A\der{y^A}-x^A\der{x^A}\right) K(X^1,X^2,X^3)=0$$ This can
be rewritten, $$\left(y^3\der{y^3}-x^3\der{x^3}
\right)K(X^1,X^2,X^3)=\left(\der{y^1}y^1-\der{x^1}x^1+\der{y^2}y^2-\der{x^2}x^2
\right)K(X^1,X^2,X^3)$$ The ``zero point'' contributions we get
from switching the order of the coordinates and derivatives on the
left fortunately cancel out. Plugging this into \EQN{kernel-form}
and integrating by parts, we immediately see that
$y\partial_y-x\partial_x$ acts as a derivation. This argument
obviously carries over to the continuum case as well. Therefore,
we may write,
$$\left(\vec{y}(\k)\cdot\frac{\delta}{\delta
\vec{y}(\k)}- \vec{x}(\k)\cdot\frac{\delta}{\delta
\vec{x}(\k)}\right)=\frac{1}{2\coth \fraction{\pi
\k}{4}}\Big[\yo(\k)\star \xe(\k)-\ye(\k)\star \xo(\k),\ \
\Big]_\star.$$ Both the left and right hand side act identically
on $1,\vec{x}(\k),\vec{y}(\k)$, and because both are derivations,
as just argued, they must act identically on all $\Psi^M$. We can
then represent the action of the reduced ghost number operator as
a star commutator,
\begin{equation}G_0(\Psi^M)=G_0^M\star \Psi^M-\Psi^M
\star G_0^M \end{equation} with,
\begin{equation}G_0^M=\half\int_0^\infty d\k\
\tanh\fraction{\pi \k}{4}\  \Big(\yo(\k)\star \xe(\k)-\ye(\k)\star
\xo(\k)\Big)
\end{equation} It is easy to see that the derivation
property of $G_0$ implies that the reduced star product is
additive in reduced ghost number, as expected.

We now turn to the BRST operator, $Q_B$, describing the background
of the unstable D-25 brane. Unfortunately, when we write it out in
the continuous oscillator basis, $Q_B$ suffers from serious
divergences. The divergences are the same as those encountered in
ref.\cite{Moyal} when they tried to formulate $L_0$ in a
continuous oscillator basis. Their presence is perhaps the most
significant shortcoming of the continuous Moyal formulation, and
it is crucial to have an explicit construction of $Q_B$ and a
concrete understanding of its divergences if this problem is to be
overcome. We are not at this point able to explicitly regulate
$Q_B$ and demonstrate that it functions properly as the regulator
is taken away, but it is still instructive to see how $Q_B$ looks
in the continuous oscillator basis. We write\cite{GSW},
\begin{eqnarray}&\ &Q_B=
\sum_{n=-\infty}^{\infty}:c_n(L_{-n}+\half L_{-n}^{gh}
-\delta_{n0}):\nonumber \\ &\ &\ \ =:c_0(L_0+\half
L_0^{gh}-1):+\int_{-\infty}^{\infty} d\k\
:\vec{\chi}(\k)\cdot\big(\vec{L}(-\k)+\half
\vec{L}^{gh}(-\k)\big):\label{eq:BRST}\end{eqnarray} The above
expression follows from the definition $\vec{\chi}(-\k) \equiv
\vec{\chi}^+(\k)$ and,\begin{eqnarray}L_\e(\k)
&=&\left\{\matrix{-i\sqrt{2}\sum_{n=1}^\infty
\frac{1}{\sqrt{2n}}v_{2n}(\k)L_{-2n}\ \ \ \k < 0 \cr
i\sqrt{2}\sum_{n=1}^\infty \frac{1}{\sqrt{2n}}v_{2n}(\k)L_{2n}\ \
\
\k > 0}\right.\nonumber\\
L_\o(\k)&=&\left\{\matrix{\sqrt{2}\sum_{n=1}^\infty
\frac{1}{\sqrt{2n-1}}v_{2n-1}(\k) L_{-2n+1}\ \ \ \k < 0 \cr
\sqrt{2}\sum_{n=1}^\infty
\frac{1}{\sqrt{2n-1}}v_{2n-1}(\k)L_{2n-1}\ \ \ \k > 0}\right.
\label{eq:cont-virasoro}\end{eqnarray} Similar expressions define
$\vec{L}^{gh}(\k)$. Let us concentrate on expressing the ghost
Virasoro generators in the continuous $bc$ oscillator basis. The
matter Virasoro generators may also be reformulated with an
appropriate choice of basis, perhaps one which diagonalizes the
relevant matter sector three string vertex. Consider first
$L_0^{gh}$,
\begin{eqnarray}L_0^{gh}&=&\sum_{n=1}^\infty n (b_n^+
c_n - b_n c_n^+)\nonumber\\ &=&\int_0^\infty d\k_1 d\k_2\
(\vec{\beta}(-\k_1)\cdot
K_{L_0}(\k_1,\k_2)\cdot\vec{\chi}(\k_2)-\vec{\beta}(\k_1)\cdot
K_{L_0}(\k_1,\k_2)\cdot\vec{\chi}(-\k_2))\nonumber\end{eqnarray}
where,
\begin{eqnarray} &\
&K_{L_0}^{\e\e}(\k_1,\k_2)=K(\k_1,\k_2)-K(\k_1,-\k_2)\nonumber\\
&\ &K_{L_0}^{\o\o}(\k_1,\k_2)=K(\k_1,\k_2)+K(\k_1,-\k_2) \nonumber\\
&\ &K_{L_0}^{\o\e}(\k_1,\k_2)=K_{L_0}^{\e\o}(\k_1,\k_2)=0
\nonumber\\\end{eqnarray} and,
\begin{equation}K(\k_1,\k_2)=\sum_{n=1}^\infty n\
v_{n}(\k_1)v_{n}(\k_2) \label{eq:div-kernel}\end{equation}
Unfortunately, the above sum does not converge, as can be seen
from the the asymptotic behavior of the $v_n$s\cite{Moyal}:
\begin{eqnarray}&\ &
v_{2n}(\k)\sim\frac{(-1)^n}{\sqrt{2n}}\frac{1}{\sqrt{
\fraction{2}{\k} \sinh \fraction{\pi
\k}{2}}}\Im\left[\frac{(4n-2)^{i\k/2}}{\Gamma(1+i\fraction{\k}{2})}\right]\nonumber
\\ &\ &
v_{2n-1}(\k)\sim\frac{(-1)^{n-1}}{\sqrt{2n-1}}\frac{1}{\sqrt{
\fraction{2}{\k} \sinh \fraction{\pi
\k}{2}}}\Re\left[\frac{(4(n-1))^{i\k/2}}{
\Gamma(1+i\fraction{\k}{2})}\right]
\label{eq:v-asymptotics}\end{eqnarray} Consider the ghost Virasoro
generators with a continuous mode label:
\begin{eqnarray}L_\e^{gh}(\k)&=&i\sqrt{2}
\sum_{m=1}^{\infty}\frac{
v_{2m}(\k)}{\sqrt{2m}}\left(\sum_{n=-\infty}^{\infty}
(2m-n)b_{2m+n}c_{-n} \right)\nonumber \\ &=& 2
b_0\ce(\k)-c_0\int_0^{\infty} d\k_1\
K_{L_0}^{\e\e}(\k_1,\k)\be(\k_1)+ \int_{-\infty}^{\infty}d\k_1
d\k_2\ \vec{\beta}(\k_1)\cdot
K_{L_\e}(\k_1,\k_2,\k)\cdot\vec{\chi}(\k_2)\nonumber\\
\label{eq:even-L}\end{eqnarray} And similarly,
\begin{eqnarray}L^{gh}_\o(\k)&=&\sqrt{2}\sum_{m=1}^{\infty}\frac{
v_{2m-1}(\k)}{\sqrt{2m-1}}\left(\sum_{n=-\infty}^{\infty}
(2m-1-n)b_{2m-1+n}c_{-n} \right)\nonumber \\ &=& 2 b_0
\co(\k)-c_0\int_0^{\infty} d\k\ K^{\o\o}_{L_0}(\k_1,\k)\bo(\k_1) +
\int_{-\infty}^{\infty}d\k_1 d\k_2\ \vec{\beta}(\k_1)\cdot
K_{L_\o}(\k_1,\k_2,\k)\cdot\vec{\chi}(\k_2) \nonumber\\
\label{eq:odd-L}\end{eqnarray} The expressions defining the
kernels $K_L$ above are somewhat lengthy, so we will not reproduce
them here. For reference and the reader's amusement, we have
written them explicitly in the appendix. They both diverge in a
similar fashion as \EQN{div-kernel}. One way to regulate these
divergences is to replace all $v_m(\k)$ with $q^m v_m(\k)$ for
$0<q<1$. However, this will almost certainly lead to heinously
complicated expressions, especially for the kernels in the
appendix which are cubic in the $v_n(\k)$s. Maybe the current
representation of the fermionic ghost star product is not the one
best suited for treatment of the BRST operator. In recent
work\cite{Bars-Matsuo} Bars and Matsuo treated the Virasoro
operators using a closely related {\it discrete} Moyal formalism
with fairly good results, though they also encounter subtleties
having to do with the fact that the Virasoro algebra doesn't close
in their regulated theory. Next section we will see how to obtain
a representation of the star product similar to theirs but in the
fermionic ghost sector. Hopefully it is possible to construct a
tractable and well-defined treatment of $Q_B$ in some formalism
where Witten's product is simple.

The divergences we're grappling with are puzzling. They must be
telling us something about our formalism, though it's unclear what
it is. Experience with split strings suggests that the midpoint
might be playing a role. We can test this hypothesis, since by
this wisdom we would expect the midpoint preserving
reparameterization generators, $\mathcal{K}_n=L_n-(-1)^n L_{-n}$
to be nonsingular. From the current standpoint this would be
remarkable coincidence, since the $\mathcal{K}_n$s are the sum of
operators which diverge. Consider for example,
\begin{eqnarray}\mathcal{K}^{gh}_1&=&
2b_0(c_1-c_1^+)+(b_1-b_1^+)c_0+\sum_{n=2}^\infty
(1+n)(b_{n-1}^+c_n-b_{n-1}c_n^+)\nonumber\\ &\ &\ \ \ \
+\sum_{n=2}^\infty(1-n) (b_{1+n}c_n^+
-b_{1+n}^+c_n)\nonumber\\
&=&-4ib_0\int_0^\infty d\k\ v_1(\k) \hat{p}_\o(\k)-2i\int_0^\infty
d\k\ v_1(\k) \hat{x}_\o(\k)\ c_0 \nonumber\\
&\ &\ -i\int_0^\infty d\k d\k'K_1(\k,\k')[\bo^+(\k)\ce(\k')+\bo(\k)\ce^+(\k')]\nonumber\\
&\ &\ +i\int_0^\infty d\k d\k'\
K_2(\k,\k')[\be^+(\k)\co(\k')+\be(\k)\co^+(\k')]\nonumber
\end{eqnarray} with
\begin{eqnarray}K_1(\k,\k')&=&2\sum_{n=1}^\infty
\left[(2n+1)\sqrt{\frac{2n-1}{2n}}v_{2n-1}(\k)v_{2n}(\k')+(2n-1)\sqrt{
\frac{2n+1}{2n}}v_{2n+1}(\k)v_{2n}(\k')\right]\nonumber
\\ K_2(\k,\k')&=&2\sum_{n=1}^\infty
\left[(2n+2)\sqrt{\frac{2n}{2n+1}}v_{2n}(\k)v_{2n+1}(\k')+2n\sqrt{
\frac{2n+2}{2n+1}}v_{2n+2}(\k)v_{2n+1}(\k')\right]\nonumber
\end{eqnarray} We see the same $n v_n^2$
terms that caused the divergences in the BRST operator, but here
these contributions actually cancel. This is because from
\EQN{v-asymptotics} we see that $v_{2n-1}$ and $v_{2n+1}$ roughly
have opposite sign for large $n$, so the first and second
divergent terms cancel. Hence $\mathcal{K}^{gh}_1$ converges. If
we had considered $\mathcal{K}^{gh}_2$ we'd have needed to compare
$v_{2n-2}$ with $v_{2n+2}$, which have the same sign, but a
cancellation still occurs in that case because we subtract the
Virasoro generators instead of adding them. In this way we can see
that all the $\mathcal{K}_n$'s (including the matter ones in the
basis of ref.\cite{Moyal}) are finite. This suggests that the
midpoint may be a part of our troubles with the BRST operator.

\section*{VII. Split string and discrete Moyal
approach to the fermionic ghosts.}

As explained in the introduction, there are at this time
essentially three proposed formalisms expressing string field
theory in the language of operator algebras and noncommutative
geometry: the split string formalism, the discrete Moyal
formalism, and the continuous Moyal formalism. So far we've mostly
developed fermionic ghosts in the continuous Moyal formalism since
this approach is the one most directly related to the three string
vertex. But clearly it is advantageous to have an understanding of
fermionic ghosts in all three formalisms, since for a particular
purpose any one might be preferable to the others.

We begin by considering the Moyal star anti-commutator of the odd
position modes and even momentum modes of $b(\sigma)$ and
$c(\sigma)$:
\begin{equation} \{y_{2m-1},q_{2n}\}_\star =
2G_{2m-1,2n},\ \ \ \
\{x_{2m-1},p_{2n}\}_\star=2\frac{2m-1}{2n}G_{2m-1,2n}
\label{eq:mode-com}\end{equation} where,
\begin{equation}G_{2m-1,2n}=-2\sqrt{\frac{2n}{2m-1}}\int_0^\infty
d\k\ \coth\fraction{\pi \k}{4}
v_{2m-1}(\k)v_{2n}(\k)\label{eq:G-matrix}\end{equation} We will
evaluate this integral by comparing it to an integral explicitly
calculated in ref.\cite{Moyal}:\begin{eqnarray}T_{2m-1,2n}&=&
-2\sqrt{\frac{2m-1}{2n}}\int_0^\infty d\k\ \tanh \fraction{\pi
\k}{4}
v_{2m-1}(\k)v_{2n}(\k) \nonumber \\
&=&\frac{2(-1)^{m+n+1}}{\pi}\left(\frac{1}{2m-1+2n}+\frac{1}{2m-1-2n}\right)
\label{eq:T-matrix}\end{eqnarray} The trick is to consider the
sum,
\begin{eqnarray}\sum_{m=1}^\infty G_{2m-1, 2n}T_{2m-1,
2n'}&=&4\int_0^\infty d\k d\k'\ \coth\fraction{\pi
\k}{4}\tanh\fraction{\pi
\k'}{4}v_{2n}(\k)v_{2n'}(\k')\sum_{m=1}^\infty
v_{2n-1}(\k)v_{2n'-1}(\k')\nonumber\\ &=&2\int_0^\infty d\k d\k'\
v_{2n}(\k)v_{2n'}(\k)\delta(\k-\k')=\delta_{2n,2n'}\nonumber
\end{eqnarray} Likewise we can argue that
$\sum_{n=1}^\infty G_{2m-1,2n}T_{2m'-1,2n}=\delta_{2m-1, 2m'-1}$.
Therefore $G$ must be the inverse of $T$. The inverse of $T$ is
known, and is a matrix often called $R$ in the literature
\cite{Bars,Bars-Matsuo,Moyal}:
\begin{equation}G_{2m-1,2n}=R_{2m-1,2n}=\frac{4n(-1)^{n+m}}{\pi
(2m-1)}\left(\frac{1}{2m-1+2n}-\frac{1}{2m-1-2n}\right).
\end{equation} A fermionic ghost version of the
discrete Moyal formalism advocated by Bars and Matsuo may be
obtained by considering a particular choice of basis,
\begin{equation} y_{2n}^D\equiv \sum_{m=1}^\infty
T_{2m-1,2n}\ y_{2m-1}\ \ \ \ x_{2n}^D\equiv \sum_{m=1}^\infty
\frac{2n}{2m-1}T_{2m-1,2n}\ x_{2m-1}
\label{eq:discrete-Moyal}\end{equation} In this basis the Moyal
star anti-commutator is particularly simple:
\begin{equation}\{y_{2m}^D,q_{2n}\}_\star=
\{x_{2m}^D,p_{2n}\}_\star=2\delta_{2m,2n}\label{eq:DM}\end{equation}
We should mention that this choice of basis encounters
difficulties having to do with the fact that both $T$ and $R$
possess zero modes and are in a sense not invertible. While it's
true that $TR=RT=1$, this is only sufficient to imply that $T$ and
$R$ are invertible when multiplication is {\it associative}, which
sometimes it isn't for such infinite dimensional
matrices\footnote{We can count on associativity only if we
restrict our space of operators, in particular to those which act
on the Hilbert space of square-integrable sequences, not the
larger Banach space of bounded sequences\cite{Moyal}.
Unfortunately, a satisfactory formulation of string field theory
seems to require the latter space.}. See \cite{associative,
Bars-Matsuo, Moyal} for further discussion of this basis and its
particular advantages and disadvantages. If we express the string
field as a functional,
$$ \Psi^D[x_{2n}^D,p_{2n},
y_{2n}^D,q_{2n}]=\Psi[x_{2n-1},y_{2n-1},p_{2n},q_{2n}]
=\Psi^M[\vec{x}(\k),\vec{y}(\k)]
$$ then in the discrete Moyal formalism one would
calculate the reduced star product as,
\begin{equation}\Psi^D*_{b_0}\Phi^D=\left.\mathcal{N}\exp\left[
\sum_{n=1}^\infty\left(\der{x_{2n}^D}\der{p_{2n}'}+\der{p_{2n}}
\der{x_{2n}^{D'}}+\der{y_{2n}^D}\der{q_{2n}'}+\der{q_{2n}}\der{y_{2n}^{D'}}\right)
\right]\Psi^{D'}\Phi^D\right|_{X=X'}\label{eq:discrete-Moyal-star}\end{equation}
This equation would be the starting point for a discrete Moyal
formulation of string field theory in the fermionic ghost sector.

Let us move on and see how to construct a split string
representation of the reduced star product. Consider as a simple
example a function of two fermionic variables, $\tilde{A}(x,p)$,
and an associated function which is given by a Fourier transform
on the $p$ variable:
\begin{eqnarray}&\ &\hat{A}(x+y,-x+y)\equiv
A(x,y)=\int dp\ e^{-py}\tilde{A}(x,p),\nonumber\\  &\
&\tilde{A}(x,p)=\int dy\
e^{py}\hat{A}(x+y,-x+y).\nonumber\end{eqnarray} It turns out that
Moyal star multiplication of functions of $x,p$ is equivalent to
calculating,
\begin{equation}\hat{A}\star\hat{B}(l,r)=-2\int dw
\hat{A}(l,-w)\hat{B}(w,r)\label{eq:dipole}\end{equation} We can
see this directly as follows:
\begin{eqnarray} \tilde{A}\star
\tilde{B}(x,p)&=&\left.e^{\der{x}\der{p'} +\der{p}\der{x'}}\int
dy' dy\ e^{p'y'}A(x'+y',-x'+y') e^{py}
B(x+y,-x+y)\right|_{X=X'}\nonumber\\
&=&\left.\int dy'dy\ e^{p'y'}e^{y\der{x'}}A(x'+y',-x'+y')
e^{-y'\der{x}}e^{py} B(x+y,-x+y)\right|_{X=X'}\nonumber\\&=&\int
dy'dy\
e^{p(y+y')}A(x+y+y',-x-y+y')B(x-y'+y,-x+y'+y)\nonumber\\
&=& -2\int dz dw\ e^{pz}A(x+z,-w)B(w,-x+z)\nonumber
\end{eqnarray} The last line follows from the
definitions $z=y+y'$ and $w=-x+y'-y$ and is easily seen to be
equivalent to \EQN{dipole}\footnote{A detailed point: above we
assumed we could commute $dy'$ through $\hat{A}$, but in the
following discussion we will actually not need this assumption
since we will always have pairs $dxdy$ for our measure, which is
Grassmann even.}. We can repeat the above argument and find a
similar representation of Moyal-star multiplication of string
fields. Define,
$$\hat{\Psi}[l_{2n}, m_{2n}, r_{2n},
s_{2n}]=\Psi[x_n,y_n]=\int\left(\prod_{m=1}^{\infty}\fraction{1}{i}
dp_{2m}dq_{2m}e^{-p_{2m} x_{2m}  -q_{2m} y_{2m}}\right)
\Psi^D[x^D_{2n}, p_{2n}, y^D_{2n}, q_{2n}]$$
$$\Psi^D[x^D_{2n}, p_{2n}, y^D_{2n},
q_{2n}]=\int\left(\prod_{m=1}^{\infty}\fraction{1}{i}
dx_{2m}dy_{2m}e^{p_{2m} x_{2m}  +q_{2m} y_{2m}
}\right)\hat{\Psi}[l_{2n}, m_{2n}, r_{2n}, s_{2n}]\nonumber $$
where,
\begin{eqnarray}l_{2m}=\fraction{1}{2m}(x_{2m}+x_{2m}^D),&\ &\
m_{2m}=y_{2m}+y_{2m}^D,\nonumber\\
r_{2m}=\fraction{1}{2m}(x_{2m}-x_{2m}^D),&\ &\
s_{2m}=y_{2m}-y_{2m}^D.\label{eq:half-modes}\end{eqnarray} Then we
can calculate the reduced star product as,
\begin{equation}\hat{\Psi}*_{b_0}\hat{\Phi}[l_{2n},
m_{2n}, r_{2n}, s_{2n}]=\mathcal{N}\int \prod_{n=1}^\infty
\fraction{8n}{i}du_{2n} dw_{2n}\ \hat{\Psi}[l_{2n}, m_{2n},
-u_{2n}, -w_{2n}]\hat{\Phi}[u_{2n}, w_{2n}, r_{2n},
s_{2n}]\label{eq:split-star}\end{equation} This equation bears a
very close resemblance to the split string formalism. Note that
``left'' and ``right'' modes are identified only up to a sign. Let
us see how this formula may be interpreted more geometrically in
terms of overlaps of $c(\sigma)$ and $b(\sigma)$. From
\EQN{half-modes} and \EQN{discrete-Moyal} we have the
correspondence,
\begin{equation}m_{2n}=y_{2n}+\sum_{m=1}^\infty
T_{2m-1,2n}y_{2m},\ \ \ s_{2n}=y_{2n}-\sum_{m=1}^\infty
T_{2m-1,2n}y_{2m-1}\label{eq:c-case}\end{equation} This equation
is easily seen to define left and right half string Fourier modes
for $c(\sigma)$ with Neumann boundary conditions at the midpoint:
\begin{eqnarray}m(\sigma)&=&m_0+\sqrt{2}\sum_{n=1}^\infty
m_{2n}\cos 2n\sigma=c(\sigma)\nonumber\\
s(\sigma)&=&s_0+\sqrt{2}\sum_{n=1}^\infty s_{2n}\cos
2n\sigma=c(\pi-\sigma)\ \ \ \ \sigma\in[0,\pi/2]
\nonumber\end{eqnarray} See ref.\cite{associative} for discussion
of split strings with Neumann boundary conditions at the midpoint.
Of course, now we see that \EQN{split-star} imposes precisely the
correct half string anti-overlap conditions on $c(\sigma)$. Note
that our string fields do not depend on the half string zero modes
$m_0$ and $s_0$ since we have taken Siegel gauge. Now let's
consider the $x$ modes,
\begin{eqnarray} l_{2n} &=& \frac{1}{2n} x_{2n}+\sum_{m=1}^\infty
T_{2m-1,2n}\frac{1}{2m-1}x_{2m-1},\nonumber\\ r_{2n} &=&
\frac{1}{2n} x_{2n}-\sum_{m=1}^\infty T_{2m-1,2n}\frac{1}{2m-1}
x_{2m-1}\nonumber\end{eqnarray} Just like \EQN{c-case} this looks
like half string Fourier modes for a field like $c(\sigma)$ but
with Fourier modes $x_n/n$:
\begin{equation}\tilde{b}(\sigma)\equiv\sqrt{2}\sum_{n=1}^\infty
\frac{x_n}{n}\cos n\sigma \end{equation} In particular, this field
satisfies $$\der{\sigma}\tilde{b}(\sigma)=i b(\sigma)$$ So we have
half string fields,
\begin{eqnarray}l(\sigma)&=&l_0+\sqrt{2}\sum_{n=1}^\infty
l_{2n}\cos 2n\sigma=\tilde{b}(\sigma)\nonumber\\
r(\sigma)&=&r_0+\sqrt{2}\sum_{n=1}^\infty r_{2n}\cos
2n\sigma=\tilde{b}(\pi-\sigma)\ \ \ \ \sigma\in[0,\pi/2]
\nonumber\end{eqnarray} Our string fields do not depend on the
half string zero modes $l_0$ and $r_0$. We can easily see that
\EQN{split-star} imposes half string anti-overlap conditions on
$\tilde{b}(\sigma)$. This is exactly correct, since setting
anti-overlap conditions on $\tilde{b}(\sigma)$ is equivalent to
setting overlap conditions on $b(\sigma)$. Note that the overlap
conditions for $\sigma\neq \frac{\pi}{2}$ should hold {\it both}
for the full and reduced star product, since they only differ by
the action of $\mathcal{Q}$, which commutes with these
constraints. Therefore we see the half string overlap conditions,
which were originally used to construct the fermionic ghost vertex
in ref.\cite{Gross-Jevicki-2}, explicitly arise out of our
construction. This is a good check of the consistency of our
results\footnote{Note: There is nothing deep about the fact that
we found Neumann boundary conditions at the midpoint in our
construction. We would have found Dirichlet boundary conditions if
we had used an odd mode discrete Moyal basis $y_{2m-1},\
q^D_{2m-1}=-\sum_{n=1}^\infty T_{2m-1,2n}q_{2n}$ and similarly for
$x,p$. Furthermore, the field $b'(\sigma)=i\partial_\sigma
b(\sigma)$ would have emerged in the split string description
rather than $\tilde{b}(\sigma)$}.

The last element we need to define Witten's product in the
discrete Moyal and split string formalisms is a legitimate
formulation of $\mathcal{Q}$. This is easy to come by. Invoking
\EQN{DM} we can write the action of $\mathcal{Q}$ on
$\Psi^D[x_{2n}^D,p_{2n}, y^D_{2n}, q_{2n}]$ as,
\begin{equation} \mathcal{Q}(\Psi^D)=c_0\Psi^D+\mathcal{Q}^D
\star\Psi^D+(-1)^{G(\Psi)}\Psi^D\star\mathcal{Q}^D\label{eq:discrete-Q}
\end{equation} where
$$\mathcal{Q}^D=\frac{1}{\sqrt{2}}\sum_{n=1}^\infty (-1)^n y_{2n}^D$$
Translating into the split string formalism, we find similarly
\begin{equation}\mathcal{Q}(\hat{\Psi})=c_0\hat{\Psi}+
\hat{\mathcal{Q}}\star\hat{\Psi}+(-1)^{G(\Psi)}\hat{\Psi}\star
\hat{\mathcal{Q}}\label{eq:split_Q}\end{equation} with
$$\hat{\mathcal{Q}}=\frac{1}{\sqrt{2}}\prod_{n=1}^\infty \frac{1}{8ni}
(l_{2n}+r_{2n})(m_{2n}+s_{2n})\sum_{n=1}^{\infty}(-1)^n m_{2n}$$
Actually, we can plug this directly into \EQN{split_Q} to find,
$$\mathcal{Q}(\hat{\Psi})=\left(c_0+\fraction{1}{\sqrt{2}}
\sum_{n=1}^\infty
(-1)^n(m_{2n}+s_{2n})\right)\hat{\Psi}=c(\fraction{\pi}{2})\hat{\Psi}
$$ This of course makes perfect sense. However, it is interesting to
note that our approach apparently fixes the way that
$c(\fraction{\pi}{2})$ must be expressed in split string variables
if it is to be identified with $\mathcal{Q}$. A priori, there is
no unique way to write $c(\fraction{\pi}{2})$ in terms of split
string variables.

\section*{IX. Conclusion}
To summarize, we have shown that the reduced star algebra of
fermionic ghosts in Siegel gauge may be formulated as a continuous
tensor product of Clifford Algebras, up to a nontrivial
normalization. We have also developed split string and discrete
Moyal representations of the reduced star product so that the
fermionic ghost sector may be studied from other viewpoints. We
have formulated the kinetic operator of VSFT and the ghost number
operator in the language of noncommutative field theory, with good
results, but the BRST operator describing the vacuum of the
unstable D-25 brane is divergent and must somehow be regulated to
get a consistent formulation.

This last point is perhaps the most crucial problem currently
facing any attempt to formulate string field theory in the
language of noncommutative geometry. After all, the only real use
of these methods is to simplify Witten's open string field theory
enough to possibly permit analytic solution to the field
equations. Vacuum string field theory, of course, seems to work
very naturally in the Moyal formulation, but to be honest VSFT is
already under comparative analytic control. Indeed, it is the
accumulation of all of this analytic understanding that makes the
results of this paper possible. Of course, we hope that our
framework might shed some light into further investigations of
VSFT, particularly it's more difficult and controversial aspects,
such as normalization of the action and the study of fluctuations
about it's various classical solutions. Still, an adequate
treatment of $Q_B$ is extremely important.

Another important open question is how one should formulate the
star product outside of Siegel gauge as a Moyal or operator
product of some sort. Siegel gauge may be useful for specific
calculations, but it would be extremely interesting to understand
the full structure of string field theory's gauge symmetry within
the operator/Moyal language. Furthermore, gauge invariance may
offer a method for generating solutions to the field equations, as
suggested for instance in ref.\cite{Kawano-Okuyama}.

Another issue is the normalization of the algebra, and whether
Witten's star product including ghost contributions truly defines
an algebra of bounded operators, as the CFT results seem to
suggest.

I would like to thank D. J. Gross for some discussions during the
development of this work. I would like to thank D. Belov for
discussion of the ghost Neumann matrices and L. Rastelli for a
conversation. This work was supported by the National Science
Foundation Grant No. PHY00-98395.

\section*{Appendix}

For reference we will write out the kernels in equations
(\ref{eq:even-L}) and (\ref{eq:odd-L}) explicitly:
\begin{eqnarray}K^{\e\e}_{L^\e}(k_1,k_2,k)&=&-4i
\sum_{m=1}^{\infty}\sum_{n=1}^{m-1}\frac{(m+n)\sqrt{m-n}}{\sqrt{mn}}
v_{2m}(k)v_{2(m-n)}(k_1)v_{2n}(k_2),\ \ \mathrm{for}\
k_1,k_2>0\nonumber\\ &=&4i
\sum_{m=1}^{\infty}\sum_{n=m+1}^{\infty}\frac{(m+n)\sqrt{n-m}}{\sqrt{mn}}
v_{2m}(k)v_{2(n-m)}(k_1)v_{2n}(k_2),\ \ \mathrm{for}\ k_1<0,\
k_2>0 \nonumber\\ &=&4i
\sum_{m=1}^{\infty}\sum_{n=1}^{\infty}\frac{(m-n)\sqrt{n+m}}{\sqrt{mn}}
v_{2m}(k)v_{2(n+m)}(k_1)v_{2n}(k_2),\ \ \mathrm{for}\ k_1>0,\
k_2<0 \nonumber\\ &=& 0 \ \ \ \mathrm{for}\ k_1,\ k_2<0\nonumber
\end{eqnarray}
\begin{eqnarray}
K^{\o\o}_{L^\e}(k_1,k_2,k)&=& i 2\sqrt{2}
\sum_{m=1}^\infty\sum_{n=1}^m\frac{(2m+2n-1)\sqrt{2m-2n+1}}{\sqrt{2m(2n-1)}}
v_{2m}(k)v_{2m-2n+1}(k_1)v_{2n-1}(k_2)\nonumber\\
\manytabs\mathrm{for}\ k_1,\ k_2>0\nonumber\\ &=& i 2\sqrt{2}
\sum_{m=1}^\infty\sum_{n=m+1}^\infty\frac{(2m+2n-1)\sqrt{2n-2m-1}}{\sqrt{2m(2n-1)}}
v_{2m}(k)v_{2n-2m-1}(k_1)v_{2n-1}(k_2)\nonumber\\
\manytabs \mathrm{for}\ k_1<0,\ k_2>0\nonumber\\ &=& i 2\sqrt{2}
\sum_{m=1}^\infty\sum_{n=1}^\infty\frac{(2m-2n+1)\sqrt{2n+2m-1}}{\sqrt{2m(2n-1)}}
v_{2m}(k)v_{2n+2m-1}(k_1)v_{2n-1}(k_2)\nonumber\\
\manytabs \mathrm{for}\ k_1>0,\ k_2<0\nonumber\\ &=& 0\ \ \
\mathrm{for}\
k_1,\ k_2<0\nonumber\\
K^{\o\e}_{L^\e}(k_1,k_2,k)&=& K^{\e\o}_{L^\e}(k_1,k_2,k)=0
\end{eqnarray}

\begin{eqnarray}
K^{\e\o}_{L^\o}(k_1,k_2,k)&=&-8
\sum_{m=1}^{\infty}\sum_{n=1}^{m-1}\frac{(m+n-1)
\sqrt{m-n}}{\sqrt{(2m-1)(2n-1)}}v_{2m-1}(k)v_{2(m-n)}(k_1)v_{2n-1}(k_2)\nonumber\\
\manytabs\mathrm{for}\ k_1,\ k_2>0 \nonumber\\ &=&8
\sum_{m=1}^{\infty}\sum_{n=m+1}^{\infty}\frac{(m+n-1)
\sqrt{n-m}}{\sqrt{(2m-1)(2n-1)}}v_{2m-1}(k)v_{2(n-m)}(k_1)v_{2n-1}(k_2)\nonumber\\
\manytabs\mathrm{for}\ k_1<0,\ k_2>0 \nonumber\\ &=&8
\sum_{m=1}^{\infty}\sum_{n=1}^{\infty}\frac{(m-n)
\sqrt{m+n-1}}{\sqrt{(2m-1)(2n-1)}}v_{2m-1}(k)v_{2(m+n-1)}(k_1)v_{2n-1}(k_2)\nonumber\\
\manytabs\mathrm{for}\ k_1>0,\ k_2<0 \nonumber\\ &=&0\ \ \
\mathrm{for}\ k_1,\ k_2<0\nonumber
\end{eqnarray}
\begin{eqnarray}
K^{\o\e}_{L^\o}(k_1,k_2,k)&=&2\sqrt{2}\sum_{m=1}^\infty
\sum_{n=1}^{m-1}\frac{(2m+2n-1)\sqrt{2m-2n-1}}{\sqrt{(2m-1)2n}}v_{2m-1}(k)v_{2m-2n-1}
(k_1)v_{2n}(k_2)\nonumber\\ \manytabs\mathrm{for}\
k_1,\ k_2>0\nonumber\\
&=&2\sqrt{2}\sum_{m=1}^\infty
\sum_{n=m}^{\infty}\frac{(2m+2n-1)\sqrt{2n-2m+1}}{\sqrt{(2m-1)2n}}v_{2m-1}(k)v_{2n-2m+1}
(k_1)v_{2n}(k_2)\nonumber\\ \manytabs\mathrm{for}\ k_1<0,\
k_2>0\nonumber\\ &=&2\sqrt{2}\sum_{m=1}^\infty
\sum_{n=1}^{\infty}\frac{(2m-2n-1)\sqrt{2n+2m-1}}{\sqrt{(2m-1)2n}}v_{2m-1}(k)v_{2n+2m-1}
(k_1)v_{2n}(k_2)\nonumber\\ \manytabs\mathrm{for}\ k_1>0,\
k_2<0\nonumber\\ &=&0\ \ \ \mathrm{for}\
k_1,k_2<0\nonumber\\
K^{\o\o}_{L^\o}(k_1,k_2,k)&=& K^{\e\e}_{L^\o}(k_1,k_2,k)=0
\end{eqnarray} Comparing the form of these equations
with \EQN{v-asymptotics}, it is clear that the above kernels all
diverge.


\begin{thebibliography}{99}

\bibitem{Sen-conjectures} A. Sen ``Universality of the
tachyon potential,'' JHEP {\bf 9912} (1999) 027 [arXiv:
hep-th/9911116].

\bibitem{Witten} E. Witten, ``Noncommutative Geometry
and String Field Theory,'' Nucl. Phys. B {\bf 268}, 253 (1986).

\bibitem{Kostelecky} V. A. Kostelecky and S. Samuel,
``The static tachyon potential in the open bosonic string
theory,'' Phys. Lett. {\bf B207} (1988) 169.

\bibitem{VSFT} L. Rastelli, A. Sen, B. Zeibach,
``String field theory around the tachyon vacuum,''
[arXiv:hep-th/0012251]

\bibitem{Sen} D. Gaiotto, L. Rastelli, A. Sen, and B.
Zwiebach, ``Ghost Structure and Closed Strings in Vacuum String
Field Theory,'' [arXiv: hep-th/0111087].

\bibitem{Gross-Jevicki-1} D. J. Gross and A. Jevicki,
``Operator Formulation of Interacting String Field Theory,'' Nucl.
Phys. B {\bf 283},1 (1987).

\bibitem{Gross-Jevicki-2} D. J. Gross and A. Jevicki,
``Operator Formulation of Interacting String Field Theory. II,''
Nucl. Phys. B {\bf 283}, 1 (1987).

\bibitem{LeClair} A. Leclair, M. E. Peskin, and C. R.
Preitschopf ``String field theory on the conformal plane, I, II,''
Nucl. Phys. B {\bf 317} 411,464, (1989).

\bibitem{half-strings} H. Chan and S. Tsun, ``String Theory considered
as a Local Gauge Theory of an Extended Object,'' Phys. Rev. D {\bf
35} (1987) 2474; J. Bordes, H. Chan, L. Nellen and S. Tsou, ``Half
String Oscillator Approach to string Field Theory,'' Nucl. Phys. B
{\bf 387} (1993) 260; L. Rastelli, A. Sen, and B. Zeibach, ``Half
strings, projectors, and multiple D-branes in vacuum string field
theory,'' JHEP {\bf 0111} (2001) 035 [arXiv:hep-th/0105058] ; D.
J. Gross and W. Taylor, ``Split String Field Theory. I,'' JHEP
{\bf 0108}, 009 (2001) [arXiv:hep-th/0105059];

\bibitem{Gross-Taylor-2} D. J. Gross and W. Taylor,
``Split String Field Theory. II,'' JHEP {\bf 0108}, 010 (2001)
[arXiv:hep-th/0106036].

\bibitem{Kawano-Okuyama} T. Kawano and K. Okuyama, ``Open String
Fields as Matrices,'' [arXiv: hep-th/0105129]

\bibitem{Bordes} A. Abdurrahman, F. Anton and J. Bordes, ``Half
string oscillator approach to string field theory (ghost sector
2),'' Nucl. Phys. B {\bf 411} (1994) 693.

\bibitem{Bars} I. Bars, ``Map of Witten's $\star$ to
Moyal's $\star$,'' Phys.Lett. {\bf B517} (2001) 436-444
[arXiv:hep-th/0106157].

\bibitem{Bars-Matsuo} I. Bars, Y. Matsuo, ``Computing
in String Field Theory using the Moyal Star Product.'' [arXiv:
hep-th/0204260].

\bibitem{Moyal} M. R. Douglas, H. Liu, G. Moore and B.
Zweibach, ``Open String Star as a Continuous Moyal Product,'' JHEP
{\bf 0204} (2002) 022 [arXiv:hep-th/0202087].

\bibitem{Aref'eva} I. Aref'eva and A. Giryavets, ``Open
Superstring Star as a continuous Moyal Product,'' [arXiv:
hep-th/0204239].

\bibitem{Belov} D. Belov, ``Diagonal Representation of
Open String Star and Moyal Product,'' [arXiv: hep-th/0204164].

\bibitem{Chen} B. Chen and F. Lin, ``D-Branes as GMS
solitons in vacuum string field theory,'' [arXiv: hep-th/0204233].

\bibitem{Douglas} M. R. Douglas and N. Nekrasov,
``Noncommutative Field theory,'' Rev. Mod. Phys. {\bf 73}, 977
(2002) [arXiv:hep-th/0106048].

\bibitem{Aref'eva_review} I. Aref'eva, D.Belov, A. Giryavets, A.
Koshelev and P. Medvedev, ``Noncommutative Field Theories and
(Super)string Field Theories.'' [arXiv:hep-th/0111208]

\bibitem{Samuel} S. Samuel, ``The physical and ghost
vertices in Witten's string field theory,'' Phys. Lett. {\bf B181}
255 (1986).

\bibitem{Spectroscopy} L. Rastelli, A. Sen, and B.
Zweibach, ``Star Algebra Spectroscopy,'' JHEP {\bf 0203} (2002)
029 [arXiv:hep-th/0111281].

\bibitem{Okuyama2} K. Okuyama, ``Siegel Gauge in
Vacuum String Field Theory,'' JHEP {\bf 0201} (2002) 043 [arXiv:
hep-th/0111087].

\bibitem{Ferrara} S. Ferrara, M.A. Lledó, ``Some
Aspects of Deformations of Supersymmetric Field Theories,'' JHEP
{\bf 0005} (2000) 008 [arXiv: hep-th/0002084]

\bibitem{GSW} M. B. Green, J. H. Schwarz, and E.
Witten, 1987, {\em Superstring theory} (Cambridge University
Press, Cambridge, England).

\bibitem{Okuyama} K. Okuyama, ``Ghost Kinetic Operator
of Vacuum String Field Theory,'' JHEP {\bf 0201} (2002) 027
[arXiv: hep-th/0201015].

\bibitem{Hata} H. Hata and T. Kuwano, ``Open String
States around a Classical Solution in Vaccuum String Field
Theory,'' JHEP {\bf 0111} (2001) {\bf 038}  [arXiv:
hep-th/0111034].

\bibitem{associative} I. Bars and Y. Matsuo,
``Associativity anomaly in string field theory,'' [arxiv:
hep-th/0202030].

\end{thebibliography}
\end{document}